\documentclass[preprintnumbers,superscriptaddress,nofootinbib,aps,prd,floatfix,twocolumn]{revtex4-2}
\pdfoutput=1
\usepackage{tikz}
\usepackage{amsmath, amssymb, amsthm, graphicx,slashed}
\usepackage{mathtools}
\usepackage[normalem]{ulem}
\usepackage{tikzsymbols}
\usepackage{natbib}
\usepackage{float}
\usepackage{amsmath}
\usepackage{amssymb}
\usepackage{amsfonts}
\usepackage{braket}
\usepackage[dvipsnames]{xcolor}
\usepackage{hyperref}
\usepackage{mathrsfs}
\hypersetup{colorlinks=true, citecolor=Blue, urlcolor=Blue, linkcolor=Blue}

\def\d{\delta}

\def\f{\frac}

\def\lam{\lambda}
\newcommand{\Trh}{T_\text{rh}}
\newcommand{\Tmax}{T_\text{max}}
\newcommand{\mdm}{m_{\rm DM}}
\newcommand{\rDR}{\rho_{\rm DR}}
\newcommand{\gs}{g_\star}
\newcommand{\gss}{g_{\star s}}

\newcommand{\be}{\begin{equation}}
\newcommand{\ee}{\end{equation}}
\newcommand{\bes}{\begin{equation*}}
\newcommand{\ees}{\end{equation*}}

\newcommand{\beq}{\begin{eqnarray}}
\newcommand{\eeq}{\end{eqnarray}}

\allowdisplaybreaks

\newcommand{\DNeff}{\Delta N_\text{eff}}
\begin{document}
\title{Baryon-dark matter coincidence in Randall-Sundrum Model}
\author{Basabendu Barman}
\email{basabendu.b@srmap.edu.in}
\affiliation{Department of Physics, School of Engineering and Sciences, SRM University AP, Amaravati 522240, India}
\author{Ashmita Das}
\email{ashmita.d@srmap.edu.in}
\affiliation{Department of Physics, School of Engineering and Sciences, SRM University AP, Amaravati 522240, India}
\author{Partha Kumar Paul}
\email{ph22resch11012@iith.ac.in}
\affiliation{Department of Physics, Indian Institute of Technology Hyderabad, Kandi, Telangana-502285, India}
\author{Narendra Sahu}
\email{nsahu@phy.iith.ac.in}
\affiliation{Department of Physics, Indian Institute of Technology Hyderabad, Kandi, Telangana-502285, India}
\author{Rakesh Kumar SivaKumar}
\email{rakesh\_sivakumar@srmap.edu.in}
\affiliation{Department of Physics, School of Engineering and Sciences, SRM University AP, Amaravati 522240, India}
\begin{abstract}
Within the framework of the extra-dimensional Randall–Sundrum set-up, we investigate the freeze-in production of Standard Model (SM) gauge-singlet scalar, fermionic, and massive vector dark matter (DM). Assuming that both the DM and SM fields reside on the IR brane and interact solely through the graviton and radion portal, we demonstrate that the Planck-observed DM relic abundance can be achieved across a wide range of reheating temperatures, all while naturally addressing the hierarchy problem, satisfying constraints from collider and early Universe cosmology. We further show that the same set-up can accommodate TeV-scale leptogenesis capable of generating the observed baryon asymmetry of the Universe. Interestingly, we find that current graviton searches at the Large Hadron Collider (LHC) already impose strong constraints on the reheating temperature in this scenario, providing a complementarity between cosmological and collider probes. 
\end{abstract}
\date{\today}
\maketitle
\preprint{\href{https://doi.org/10.1103/qjwc-snqq}{10.1103/qjwc-snqq}}
\noindent
\section{Introduction}
\label{sec:intro}
Despite the overwhelming astrophysical and cosmological evidence for the existence of dark matter (DM)~\cite{Bertone:2016nfn, deSwart:2017heh}, its particle nature and its interactions with the Standard Model (SM) remain among the most profound open questions. To date, the presence of this elusive component has only been inferred from its gravitational effects. It is conceivable that DM interacts solely through gravity and remains completely inaccessible to present or future particle physics experiments. In such a case, reproducing the observed DM relic abundance purely through gravitational production would require a very high reheating temperature~\cite{Ema:2015dka,Garny:2015sjg,Tang:2016vch,Ema:2016hlw,Garny:2017kha,Tang:2017hvq,Bernal:2018qlk,Kolb:2023ydq} because the strength of gravity is set by the large Planck mass. This inference, however, applies only if the Universe is strictly four-dimensional. In the presence of extra dimensions, gravitational interactions can be significantly enhanced at short distance either because the fundamental Planck scale in $D$ dimensions is much lower than $M_P$ as in the case of Large Extra Dimensions (LED)~\cite{Antoniadis:1990ew,Antoniadis:1997zg,ArkaniHamed:1998rs,Antoniadis:1998ig}, or due to a warped geometry that induces a much smaller effective coupling on our four-dimensional brane, as in the Randall--Sundrum (RS) models~\cite{Randall:1999ee,Randall:1999vf}. These features have long been invoked as possible solutions to the hierarchy problem, which refers to the puzzling gap between the electroweak scale and the Planck scale, whose large value would otherwise induce Planck-scale corrections to the Higgs mass. Without an enormous amount of fine-tuning or the unlikely assumption that the SM is the ultimate theory, these corrections would destabilize the observed Higgs mass. In such models, either lowering the effective Planck scale or via a direct warping of the bare mass parameters on the $(3+1)$ dimensional brane, the Higgs mass hierarchy problem can be resolved. Furthermore in these models, the effective coupling of the matter fields with the higher excited Kaluza-Klein modes of graviton acquire an enhancement with respect to the massless graviton. This in turn reinforces observational prominence of these models in the ongoing collider experiments.  As a result, a DM particle interacting only through gravity could behave like a weakly interacting massive particle (WIMP) with a typical mass in the $100$--$1000\,\mathrm{GeV}$ range and a relic abundance set by thermal freeze-out. This possibility has been extensively studied within the RS models~\cite{Lee:2013bua,Lee:2014caa,Han:2015cty,Rueter:2017nbk,Rizzo:2018ntg,Carrillo-Monteverde:2018phy,Rizzo:2018joy,Brax:2019koq,Folgado:2019sgz,Chivukula:2025pmk} as well as in several recent works investigating generic spin-2 mediators~\cite{Kang:2020huh,Chivukula:2020hvi,Kang:2020yul,Kang:2020afi}\footnote{For completeness, let us also mention the recent studies on DM {\it freeze-out} in the context of RS model with 3-branes, performed in, for example, Refs.~\cite{Koutroulis:2024wjl,Donini:2025cpl}.}. However, recent advances in WIMP search experiments have placed severe constraints on the WIMP paradigm~\cite{Arcadi:2017kky,Roszkowski:2017nbc}, thereby motivating alternative mechanisms for DM production. One well-studied alternative is the feebly interacting massive particle (FIMP) framework, in which DM is produced from the decay or annihilation of visible-sector particles in the early Universe. As the temperature of the SM plasma (the radiation bath) drops below the relevant interaction scale, DM production becomes Boltzmann suppressed, leading to a constant comoving number density. This process is known as freeze-in~\cite{McDonald:2001vt,McDonald:2008ua,Hall:2009bx,Bernal:2017kxu}. The FIMP framework requires extremely suppressed interactions between the dark and visible sectors to ensure non-thermal production. Such interactions can originate from either small infrared couplings or higher-dimensional non-renormalizable operators suppressed by a large new-physics (NP) scale, as in the ultraviolet (UV) freeze-in mechanism~\cite{Hall:2009bx,Elahi:2014fsa}.

On the other hand, observations from two independent cosmological probes, the cosmic microwave background (CMB)~\cite{Planck:2018vyg} and big bang nucleosynthesis (BBN)~\cite{Kolb:1990vq,Steigman:2005uz,ParticleDataGroup:2006fqo}, confirm that the visible (baryonic) matter in the Universe is asymmetric. Explaining this baryon asymmetry poses a fundamental challenge in particle physics. According to Sakharov’s conditions~\cite{Sakharov:1967dj}, three key ingredients are required for the dynamical generation of such an asymmetry: (i) baryon number violation, (ii) violation of $C$ and $CP$ symmetries, and (iii) departure from thermal equilibrium. Although the SM contains these elements in principle, they are insufficient to account for the observed magnitude of the asymmetry, motivating the need for new physics beyond the SM. A particularly compelling explanation is leptogenesis~\cite{Fukugita:1986hr}, which generates the baryon asymmetry of the Universe (BAU) via the asymmetry generated in the lepton sector. In this framework, a lepton asymmetry is first produced and subsequently converted into a baryon asymmetry through $(B+L)$-violating electroweak sphaleron transitions~\cite{Kuzmin:1985mm}. When implemented in the Type-I seesaw mechanism—originally proposed to explain the origin of neutrino masses~\cite{Minkowski:1977sc,GellMann:1980vs,Mohapatra:1979ia,Schechter:1980gr,Schechter:1981cv}—the complex Yukawa couplings of heavy right-handed neutrinos (RHNs) with SM leptons and Higgs doublets provide the required $CP$ violation. The expansion of the Universe (quantified by the Hubble rate) drives the RHN decays out of thermal equilibrium, while their Majorana masses violate lepton number. In the standard thermal leptogenesis scenario (for a review, see Ref.~\cite{Davidson:2008bu}), RHNs are thermally produced from the SM plasma. A lower bound on their masses, known as the Davidson--Ibarra bound, imposes a constraint on the reheating temperature, requiring $\Trh \gtrsim 10^9~\text{GeV}$~\cite{Davidson:2002qv}. Once produced, their out-of-equilibrium decays can successfully generate the observed matter--antimatter asymmetry of the Universe.
\begin{figure}
    \centering
    \includegraphics[scale=0.085]{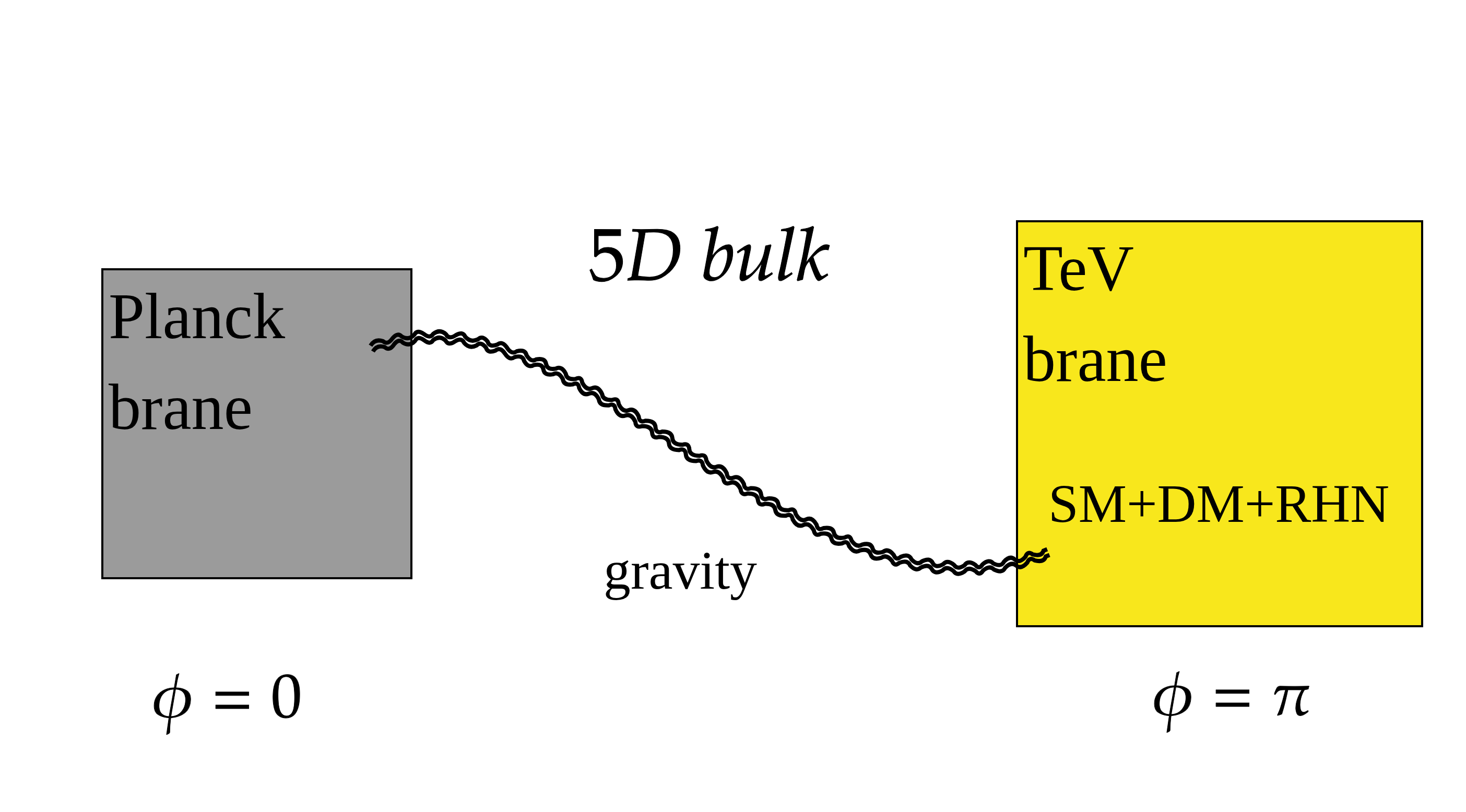}
    \caption{Schematic diagram showing our scenario. Only graviton is capable of accessing the bulk. This figure is generated using online math editor {\texttt mathcha.io}.}
    \label{fig:RS}
\end{figure}

Motivated from these, in this work we explore a framework where both DM and the BAU are generated in the context of a warped RS geometry (Fig.~\ref{fig:RS} shows a schematic). To ensure that the hierarchy problem remains addressed, we restrict the DM mass to the TeV scale and consider three possible spin assignments for the DM candidate: a scalar, a Majorana fermion, or a massive gauge boson. Depending on the DM mass, the observed relic abundance can be achieved for reheating temperatures as low as $\Trh \lesssim 10^5~\text{GeV}$, while remaining consistent with existing cosmological and collider constraints. In the same framework, the production of right-handed neutrinos (RHNs) via scatterings of thermal bath particles (via radion and graviton mediation) can generate the required matter--antimatter asymmetry through their CP-violating out of equilibrium decay. Remarkably, the warping-induced corrections to their Majorana masses naturally enable leptogenesis to occur at the TeV scale. Overall, this framework provides a TeV-scale leptogenesis scenario that simultaneously explains observed DM abundance, and points toward the existence of warped extra dimensions. We would like to highlight that the present study differs from earlier works in several ways. For instance, in~\cite{Bernal:2020fvw}, the resolution of the hierarchy problem plays a comparatively secondary role, while the authors mostly focus on DM production mechanisms. In contrast, we treat the hierarchy issue as central and delineate the region of parameter space where both the hierarchy problem and either the correct DM relic density or successful baryogenesis are simultaneously addressed. Moreover, Ref.~\cite{Bernal:2020fvw} does not examine the implications for leptogenesis, which we incorporate in our analysis. Freeze-in production of DM via KK-graviton (in the context of both Large Extra Dimensions (LED) and warped extra dimensions in the RS model) decay has been studied in Ref.~\cite{deGiorgi:2022yha}, where the focus lies on generating warm DM from the decay of heavy, long-lived KK modes. In contrast, in our framework the KK-gravitons act primarily as {\it portals} mediating interactions between the SM and the DM sector. 

The paper is organized as follows. We present the model set-up in Sec.~\ref{sec:framework}. The dynamics of DM production is elaborated in Sec.~\ref{sec:freezein}, followed by the discussion on baryogenesis via leptogenesis in Sec.~\ref{sec:lepto}. Finally, we conclude in Sec.~\ref{sec:concl}.
\section{The set-up}
\label{sec:framework}
We consider the {\it minimal} RS setup following~\cite{Randall:1999ee,Randall:1999vf}. This involves $S^1/Z_2$ orbifolding with two 3-branes at $\phi=0$ and $\phi=\pi$. The bulk cosmological constant and brane tensions generate a non-factorizable metric,  
\begin{equation}
\label{eq:metric}
ds^2 = e^{-2 k r_c |\phi|}\eta_{\mu\nu} dx^\mu dx^\nu - r_c^2 d\phi^2\,,
\end{equation}
where $k \sim M$ and $r_c$ sets the size of the fifth dimension. The effective four-dimensional Planck scale is,  
\begin{equation}
M_P^2=\frac{M^3}{k}\left(1-e^{-2 k r_c \pi}\right)\,,
\end{equation}
where $M$ is the fundamental Planck scale. For $k r_c \simeq 11-12$, this generates the weak scale dynamically from a fundamental Planckian scale via the warp factor $e^{-kr_c\pi}$. A generic prediction of the model is TeV-scale Kaluza–Klein excitations of bulk fields, including the graviton, with couplings suppressed at the TeV scale.
\subsection{Introducing the radion}
\label{radion}
In this section, we briefly review the origin of the radion in the RS framework. In the original RS setup, a massless scalar degree of freedom naturally appears in the 4-dimensional effective theory. This scalar, known as the radion, corresponds to fluctuations of the compactification radius $r_c$ in Eq.~\eqref{eq:metric}, and couples to matter with TeV-scale strength. Allowing the size of the extra dimension to fluctuate, the RS metric can be written as  
\begin{equation}
\label{met_fld_fluct}
ds^2 = e^{-2 k T(x) |\phi|}g_{\mu\nu}(x)\, dx^\mu dx^\nu - T^2(x)\, d\phi^2\,,
\end{equation}
where the fixed radius $r_c$ is promoted to a dynamical field $T(x)$, with vacuum expectation value $\langle T \rangle = r_c$~\cite{Goldberger:1999un}. Defining 
\begin{align}
\varphi=f\,{\rm exp}(-k\pi T), \qquad f=\sqrt{\frac{24 M^3}{k}}\,,
\end{align}
it was shown in Ref.~\cite{Goldberger:1999un} that the original RS model does not naturally generate a stabilizing potential for this modulus field. To stabilize the extra dimensional radius, Goldberger and Wise introduced a massive bulk scalar field $\Phi$ with mass $m$, along with localized potentials on the hidden and visible branes characterized by boundary values $v_{\rm hid}$ and $v_{\rm vis}$, respectively~\cite{Goldberger:1999uk}. This mechanism generates an effective potential for the 4-dimensional radion field:  
\begin{equation}
\label{radion_pot_1}
V(\varphi)=\frac{k^3\,\varphi}{144 M^6}\left[v_{\rm vis}-v_{\rm hid}\left(\varphi/f\right)^{\epsilon}\right]^2,
\qquad
\epsilon \equiv \frac{m^2}{4\,k^2}\ll 1\,,
\end{equation}
where terms of order $\mathcal{O}(\epsilon)$ have been neglected~\cite{Goldberger:1999un}. For a detailed discussion of the stabilization mechanism, we refer the reader to Ref.~\cite{Goldberger:1999uk}.  

A few comments are useful here regarding the low-energy scalar spectrum. In a generic 5-dimensional Kaluza--Klein (KK) theory, the metric decomposes into a 4-dimensional graviton $g_{\mu\nu}^{(4)}$, a graviphoton $A_\mu^{(4)}$, and a graviscalar $\varphi$, corresponding respectively to the tensor, vector, and scalar components of the higher-dimensional metric. In the stabilized RS framework, the Goldberger--Wise scalar $\Phi$ mixes with the graviscalar mode $g_{44}^{(5)}$. However, the KK graviphoton modes become massive by absorbing the graviscalar KK tower, removing these scalar KK excitations from the low-energy spectrum. Meanwhile, the KK excitations of the stabilizing field $\Phi$ are typically much heavier and therefore phenomenologically irrelevant at low energies. As a result, the low-energy theory contains a single light scalar state arising from the combination of the zeroth modes of $\Phi$ and the graviscalar. Expanding the radion field around its vacuum expectation value,  
\begin{align}
\varphi=\langle\varphi\rangle+\delta\varphi \equiv \Lambda_r+r(x)\,,
\end{align}
one identifies $r(x)$ as the physical radion field, while $\Lambda_r \equiv \langle\varphi\rangle$ sets the scale of radion interactions. The radion mass is then obtained from Eq.~\eqref{radion_pot_1}:  
\begin{align}
m_r^2 =
k^2 v_{\rm vis}^2
\left(
\frac{2\,\epsilon^2\,e^{-2\pi k r_c}}{3 M^3}
\right)\,.
\end{align}
An important feature is that the radion can naturally be much lighter than the first KK graviton excitation. We also note that both the bulk scalar field $\Phi$ and the brane parameter $v_{\rm vis}$ carry mass dimension 3/2.

\subsection{Radion interactions with the SM}
\label{sec:rad-SM}
We assume that the SM fields are located on the 4D brane. Therefore to start with, we consider the 5-dimensional gravitational action together with the $(3+1)$ dimensional brane action in the RS scenario. We derive the corresponding  effective 4-dimensional action, allowing up to the 1st order of the perturbative expansion of $\varphi$ as follows,
\begin{eqnarray}\label{eq:radion-SM-action}
& \mathcal{S}_{\rm rad-SM}\supset\int d^4x \sqrt{-g}\,\left(\f{4\,r}{\sqrt{6}\,\Lambda_r}\right)\biggl[-\,\f{1}{2}\left(D_{\mu}H\right)^{\dagger}\left(D^{\mu}H\right)+\nonumber\\
&\mu^{2}\,H^{\dagger} H-\lambda \left(H^{\dagger} H\right)^2-m_{\psi}\bar{\psi}\,\psi-\nonumber\\
&\f{y_\psi}{\sqrt{2}}\biggl\{\bar{\psi}_L\,h(x)\,\psi_R+\bar{\psi}_R\,h(x)\,\psi_L\biggr\}\biggr]\,.
\end{eqnarray}
In the above, $H=\left(0~~~~h/\sqrt{2}\right)^T$ is the SM Higgs field in the unitary gauge, $\psi$'s are the SM fermions, with $L(R)$ representing their left (right) chirality. The covariant derivative is defined as: $D_\mu=\partial_\mu-\left(i\,g_2/2\right)\,\sigma^a\,W_\mu^a-\left(i\,g_1/2\right)\,B_\mu$, where $g_{2(1)}$ are the gauge couplings along with the gauge bosons $W(B)_\mu$'s corresponding to $SU(2)_L$ and $U(1)_Y$, respectively, and $\sigma$'s are the Pauli spin matrices. Further, the induced metric on the visible and hidden brane can be written as, $g_{\mu\nu}^{\rm vis}(x^{\mu})\equiv e^{-k T(x)\pi}g_{\mu\nu}$ and  $g_{\mu\nu}^{\rm hid}(x^{\mu})\equiv g_{\mu\nu}$, respectively. On the visible brane, 
\begin{align}
\Lambda_r=M_P\,e^{-kr_c\,\pi}\,,   
\end{align}
with the redefined Higgs mass as,
\begin{align}\label{eq:mass-warp}
\mu\to\mu_0\, e^{-k r_c \pi}\,, 
\end{align}
to ensure that the kinetic term remains canonical. For $\mu_0\sim\mathcal{O}(M_P)$ and $kr_c\sim\mathcal{O}(10)$, the physical mass $\mu$ is of the order of weak scale, thereby solving the issue of gauge hierarchy problem, which refers to the vast disparity between the weak scale and the Planck scale. This also implies, for $kr_c\sim 10$, $\Lambda_r\sim 10$ TeV. The kinetic sector of the gauge field has no interaction with radion. One can write the action in Eq.~\eqref{eq:radion-SM-action} in a more compact form~\cite{Csaki:1999mp,Kribs:2001ic,Csaki:2000zn},
\begin{align}
& \mathcal{S}_{\rm rad-SM}\supset\int d^4x \sqrt{-g}\,\left(\frac{r}{\sqrt{6}\,\Lambda_r}\right)\,T_\alpha^\alpha\Big|_{\rm SM}\,,    
\end{align}
where $T_\alpha^\alpha\big|_{\rm SM}$ denotes the trace of the energy momentum tensor for all the SM fields\footnote{It is worth noting that while the radion field couples to both SM and DM particles through the trace of the energy–momentum tensor, photons and gluons are an exception, as being massless they do not contribute to the trace at tree level. Nevertheless, effective couplings to the radion can arise via quark and $W$-boson loops as well as through the trace anomaly~\cite{Blum:2014jca,Csaki:1999mp,Folgado:2019sgz}.}. 
\subsection{Radion interactions with bosonic DM}
\label{sec:rad-dm}
In order to explain the DM abundance, we extend the SM particle content by first adding a singlet scalar $S$, odd under some ad-hoc $Z_2$ symmetry that ensures its stability. We consider, for now, the DM fields are located on the 4D brane, similar to the SM fields. Thus the interaction Lagrangian can be written as follows,
\begin{align}
&\mathcal{L}_S\supset\frac{1}{2}\,\partial_\mu S\,\partial^\mu S-\frac{1}{2}\,m_S^2\,S^2-\frac{\lam_p}{2}S^2|H|^2-\frac{\lam_S}{4}S^4\,, 
\end{align}
where the term proportional to $\lambda_p$ corresponds to the Higgs portal interaction, while the last term gives rise to DM self-interaction. In the following analysis we will consider the Higgs portal interaction term $S^2\,|H|^2$ and the self-interactions $S^4$ terms are absent by construction. Under such conditions and starting from the action of the singlet $S$ on the $(3+1)$ dimensional brane, we find the action corresponding to the interaction of the DM $S$ and the radion as below,
\begin{align}\label{eq:rad-scaldm}
\mathcal{S}_{{\rm rad}-S}\supset\int d^4x \sqrt{-g}\,\f{r}{\sqrt{6}\,\Lambda_r}\,\left[-\,\partial_{\mu}S\,\partial^{\mu}S+\,2\,m_S^{2}\,S^{2}\right]\,.
\end{align}
Note that, even in the absence of the Higgs portal, the DM can still interact with the visible sector, thanks to the radion mediation.

Similar to the case of scalar DM, one can imagine a massive spin-1 DM $X_\mu$ with the following Lagrangian,
\begin{align}
&\mathcal{L}_X\supset-\frac{1}{4}\,X_{\mu\nu}\,X^{\mu\nu}+\frac{1}{2}\,m_X^2\,X_\mu\,X^\mu\,,   
\end{align}
where the first term is the usual gauge boson kinetic term, while the second is a bare mass term. Here we assume a Stueckelberg mass\footnote{In abelian gauge theories, the Stueckelberg mechanism can be taken as the limit of the Higgs mechanism where the mass of the real scalar is sent to infinity and only the pseudoscalar is present~\cite{Stueckelberg:1938hvi, Ruegg:2003ps, Kors:2004dx, Kors:2005uz}.} term for the new gauge boson $X_\mu$. We also assume the presence of an unbroken $Z_2$ symmetry under which the DM is odd while all the SM fields are even, thus ensuring the stability of the DM by forbidding the kinetic mixing term. With this, one can write down the action encoding interaction between $X_\mu$ and radion,
\begin{align}\label{eq:rad-vecdm}
& \mathcal{S}_{{\rm rad}-X_\mu}\supset\int d^4x \sqrt{-g}\,\f{r}{\sqrt{6}\,\Lambda_r}\,\left(\frac{m_X^2}{2}\,X_\mu\,X^\mu\right)\,.
\end{align}
Both $m_S$ and $m_X$ are scaled by the RS warp factor, similar to the Higgs mass in Eq.~\eqref{eq:mass-warp}. It is worth mentioning that we consider each DM is present one at a time and not together. Once again, we have redefined the masses of $S$ and $X_\mu$ as,
\begin{align}\label{eq:mdm1}
& m_S\to m_S^0\,e^{-kr_c\,\pi}\,, & m_X\to m_X^0\,e^{-kr_c\,\pi}\,,    
\end{align}
which shows, addressing the gauge hierarchy issue, in the present framework, one can realize DM in the ballpark of 100 GeV to 1 TeV for $kr_c\simeq\mathcal{O}(10)$ and $m_{S,X}^0\sim\mathcal{O}(M_P)$.
\subsection{Radion interactions with fermionic DM}
\label{sec:rad-RHN}
We next consider a singlet fermionic Majorana DM candidate $\Psi$, protected by an ad-hoc $Z_2$ symmetry, with the following interaction Lagrangian,
\begin{equation} \label{eq:chi-lgrng}
\mathcal{L} \supset i\,\overline{\Psi}\,\slashed{\nabla}\Psi-\frac12\, m_\Psi\, \overline{\Psi^c}\, \Psi+{\rm H.c.}\,,
\end{equation}   
where $m_\Psi$ is the Majorana mass. 

In case of fermion the covariant derivative is defined as: $\nabla_\mu\,\Psi=\,\partial_\mu\,\Psi+\frac{i}{2}\,\omega_{\mu}\,^{mn}\,\sigma_{mn}\,\Psi$, where $\omega_{\mu}\,^{mn}$ are the antisymmetric coefficients of the spinor connection and $\sigma_{mn}=\,\frac{i}{2}\,(\gamma_{m}\gamma_{n}-\gamma_{n}\gamma_{m})$.

We write, $\omega_{\mu}\,^{mn}=\,e_{\nu}^{m}\,\Gamma^{\nu}_{\sigma\mu}\,e^{\sigma  n}-\,(\partial_{\mu}\,e^{m}_{\nu})\,e^{\nu n}$, where $e_{\nu}^{m}=e^{k\pi T(x)}\d_{\nu}^{m}$ and $e_{m}^{\nu}=e^{-k\pi T(x)}\d_{m}^{\nu}$, are called vierbeins (see Appendix of \cite{Barman:2021qds} for a detailed derivation). With the Lagrangian in Eq.~\eqref{eq:chi-lgrng}, we can write down the subsequent action as,
\begin{align}
\mathcal{S}_{\text{rad}-\Psi}\supset\,-2\,m_{\Psi}\int d^4x\,\left(\f{r}{\sqrt{6}\,\Lambda_r}\right) \,\overline{\left(\Psi^c\right)}\, \Psi\,.
\end{align}
It is worth noting that here we have rescaled the mass as
\begin{align}\label{eq:mdm2}
m_\Psi\to m_\Psi^0\,e^{-k\,r_c\,\pi}\,.    
\end{align} 
\subsection{Graviton-matter interactions}
\label{sec:grav-matter}
In this section we study fluctuations of the graviton field by $h_{\mu\nu}(x,\phi)$, treated as the perturbation around the flat-brane metric $\eta_{\mu\nu}$ such as, $\hat{g}_{\alpha\beta}=e^{-2\sigma}\left(\eta_{\mu\nu}+\kappa^{*}h_{\mu\nu}\right)$, where $\sigma(\phi)\equiv kr_c\,|\phi|$. Here $\kappa^{*}$ represents the expansion parameter. In terms of the KK mode decomposition $h_{\mu\nu}(x,\phi)$ can be written as follows \cite{Davoudiasl:1999jd}, 
\begin{align}
& h_{\mu\nu}(x,\,\phi)=\sum_n h_{\mu\nu}^{(n)}(x)\,\frac{\chi^{(n)}(\phi)}{\sqrt{r_c}}\,,    
\end{align}
where $h_{\mu\nu}^{(n)}(x)$ can be interpreted as the KK-excitations of the 4-dimensional graviton, and $\chi^{(n)}(\phi)$ are the wavefunctions of the KK gravitons along the extra dimension. To extract the KK graviton mass modes, we rewrite the 4-dimensional Einstein's equation with respect to the perturbed metric $\hat{g}_{\alpha\beta}$. Allowing upto the linear order of $\kappa^{*}$ and imposing the gauge conditions: $\partial^{\alpha}h_{\alpha\beta}=h_{\alpha}^{\alpha}=0$, we obtain the equation of motion for $h_{\mu\nu}$ as follows, 
\begin{equation}
\left(\eta^{\mu\nu} \partial_\mu\,\partial_\nu + m_{G_n}^2\right) 
h_{\mu\nu}^{(n)}(x)=0\,,
\label{eom}
\end{equation}
corresponding to states with masses $m_{G_n} \geq 0$. Employing the KK expansion in the metric $\hat{g}_{\alpha \beta}$, 
Einstein’s equation together with the above relation yields a differential equation for the profile $\chi^{(n)}(\phi)$,
\begin{equation}
-\frac{1}{r_c^2} \frac{d}{d \phi} \left(e^{- 4 \sigma} \frac{d \chi^{(n)}}{d 
\phi}\right) = m_{G_n}^2 \, e^{- 2 \sigma} \chi^{(n)}\,.       
\label{eq:diffeq}
\end{equation}
The solution for $\chi^{(n)}$ reads~\cite{Goldberger:1999wh,Davoudiasl:1999jd},
\begin{equation}
\chi^{(n)}(\phi) = \frac{e^{2 \sigma}}{N_n} \Big[J_2 (z_n) + \alpha_n \, Y_2 
(z_n)\Big]\,, 
\label{eq:chi}
\end{equation}
along with the orthonormality condition 
\begin{align}
\int_{-\pi}^{\pi} d\phi \, e^{-2\sigma}\,\chi^{(m)}(\phi)\chi^{(n)}(\phi)
= \delta_{mn}\,,    
\end{align}
where $J_2$ and $Y_2$ denote Bessel functions of order 2, $z_n (\phi) = m_{G_n} e^{\sigma (\phi)}/k$, $N_n$ is the normalization constant, and $\alpha_n$ are fixed coefficients. The KK graviton masses are then
\begin{align}\label{eq:mn}
m_{G_n} = k x_n e^{-kr_c\pi}\,.    
\end{align}
The normalization factors can be computed demanding,
\begin{align}\label{eq:norm-int}
& 2\,\int_0^\pi\,d\phi\,e^{-2\sigma(\phi)}\,\left(\chi^{(n)}(\phi)\right)^2=1\,.    
\end{align}
Note that, there is a net $e^{2\sigma(\phi)}$ sitting inside the integral, which makes the normalization factors $\gtrsim\mathcal{O}(10^{10})$. For $x_n \ll e^{kr_c\pi}$, where $x_n\equiv z_n(\phi=\pi)$, one finds $\alpha_n \ll 1$, allowing the $Y_2(z_n)$ term in Eq.~\eqref{eq:chi} to be 
neglected. Further, the
requirement that the first order derivative of $\chi^{(n)}$ be continuous at the orbifold fixed points yields,
\begin{align}
& \alpha_n=-\frac{J_1\left(m_{G_n}/k\right)}{Y_1\left(m_{G_n}/k\right)}\,, & J_1(x_n)=0\,.    
\end{align}
The normalization constants are then
\begin{align}
& N_n \simeq \frac{e^{kr_c\pi}}{\sqrt{kr_c}}\, J_2(x_n)\,, & n>0\,,
\label{Nn}
\end{align}
and the zero mode is normalized as $N_0 = 1/\sqrt{kr_c}$. We note that Ref.~\cite{Folgado:2019sgz} obtains the zero-mode normalization constant $N_0$ with an overall negative sign. This is consistent with Eq.~\eqref{eq:norm-int}, which fixes only $N_0^2$, giving
\begin{align*}
N_0 = \pm \sqrt{\frac{1-e^{-2kr_c\pi}}{kr_c}}
\;\simeq\;
\pm \frac{1}{\sqrt{kr_c}}\,;
\qquad (kr_c\pi \gg 1)\,.
\end{align*}
These two choices are physically equivalent, since they differ only by an overall sign redefinition of the graviton zero-mode wavefunction (or equivalently the corresponding 4-dimensional graviton field), which leaves measurable quantities such as decay widths and scattering cross sections unchanged. In Tab.~\ref{tab:mn} we have listed KK graviton masses for $n=1,...,10$, considering $kr_c=11$ and $k=M_P$. It is worth noting that the spectrum of graviton masses is highly sensitive to the value of $k$. Here we have fixed $k$ to be at the Planck scale. However, if $k$ is taken to be slightly below the Planck scale, the mass of the first graviton KK mode can be as low as $\sim 2$ TeV, placing it within the reach of collider experiments, as we discuss in subsection~\ref{sec:constraints}. For now, however, our focus is not on collider search prospects, but rather on the potential cosmological implications for DM production, and hence we choose the graviton mass spectrum well beyond the collider bounds. 

At large values of $n$, the roots
of the Bessel function become approximately $x_n\simeq \pi\,\left(n+1/4\right)+\mathcal{O}(n^{-1})$.
\begin{table*}[htb!]
\centering
\small
\renewcommand{\arraystretch}{1.2}
\setlength{\tabcolsep}{8pt}
\begin{tabular}{c | c c c c c c c c c c c c} 
\hline 
$n$ & 1 & 2 & 3 & 4 & 5 & 6 & 7 & 8 & 9 & 10\\
\hline\hline
$m_{G_n}$ [TeV] &  9.15 & 16.76 & 24.31 & 31.84 & 39.36 & 46.87 & 54.39 & 61.90 & 69.41 & 76.92\\
\hline
\end{tabular}
\caption{Graviton masses corresponding to a few KK modes $(n=1,\,...,10)$ for $kr_c=11$ and $k=M_P$, calculated using Eq.~\eqref{eq:mn}.} 
\label{tab:mn}
\end{table*}
With the solutions for $\chi^{(n)}$ in hand, the interaction of 
$h_{\mu\nu}^{(n)}$ with brane-localized matter fields can now be derived. 
Starting from the 5D action and restricting to the brane at $\phi=\pi$, one obtains
\begin{align}
\mathcal{L}=-\frac{1}{M^{3/2}}\, T^{\mu\nu}(x)\, h_{\mu\nu}(x,\phi=\pi)\,,  
\end{align}
where $T^{\mu\nu}(x)$ is the conserved 4D energy–momentum tensor of matter 
fields. Expanding $h_{\mu\nu}$ in KK modes and using 
Eq.~\eqref{Nn} for normalization, one finds
\begin{equation}\label{eq:Lgraviton}
\mathcal{L} = - \frac{1}{M_P}\,T^{\mu\nu}(x)\,h^{(0)}_{\mu\nu}(x)
- \frac{1}{\Lambda_r}\, T^{\mu\nu}(x)\sum_{n=1}^\infty h^{(n)}_{\mu\nu}(x)\,,
\end{equation}
where the zero mode couples with the usual 4D gravitational strength 
$M_P^{-1}$. 

Now, the total energy–momentum tensor appearing in Eq.~\eqref{eq:Lgraviton} can be expressed as  
\begin{align}
    T^{\mu\nu} = T^{\mu\nu}_{\rm SM} + T^{\mu\nu}_{\rm DM}\,,
\end{align}
where the SM contribution is given by,  
\begin{align}
    T_{\mu\nu}^{\rm SM} = T_{\mu\nu}^\psi + T_{\mu\nu}^H + T_{\mu\nu}^V\,.
\end{align}
Explicitly, the individual components corresponding to the SM fermions, Higgs scalar, and gauge bosons take the forms  
\begin{eqnarray}
    T_{\mu\nu}^\psi &=& i\left[\bar{\psi}\,\gamma_{\nu}\,\partial_{\mu}\psi
    - \left(\partial_{\mu}\bar{\psi}\right)\gamma_{\nu}\psi\right]
    - \frac{i}{2}\,\eta_{\mu\nu}
    \left[\bar{\psi}\,\gamma^{\alpha}\,\partial_{\alpha}\,\psi
     \right.\nonumber\\&&-\left.\left(\partial_{\alpha}\bar{\psi}\right)\gamma^{\alpha}\psi\right],
    \nonumber\\
    T_{\mu\nu}^H &=& \left(D_{\mu}H\right)^{\dagger}\left(D_{\nu}H\right)
    + \left(D_{\nu}H\right)^{\dagger}\left(D_{\mu}H\right)
    -\nonumber\\&& \eta_{\mu\nu}\!\left[
    \left(D_{\alpha}H\right)^{\dagger}\left(D^{\alpha}H\right)
    + \mu^{2}\,H^{\dagger}\,H
    - \lambda\left(H^{\dagger}\,H\right)^{2}\right],
    \nonumber\\
    T_{\mu\nu}^V &=& -\,\eta^{\alpha\beta}\,V_{\nu\beta}^{(a)}\,V_{\mu\alpha}^{(a)}
    + \frac{1}{4}\,\eta_{\mu\nu}\,V_{\alpha\beta}^{(a)}\,V^{(a)\alpha\beta},
    ~ (a=1,2,3)\,\nonumber\\
\end{eqnarray}
The canonical energy–momentum tensors corresponding to the DM fields with spins \( i = 0,\, \tfrac{1}{2},\, 1 \) are given by, 
\begin{eqnarray}
    T^{\mu\nu}_S &=&
    \partial^{\mu}S\,\partial^{\nu}S
    - \eta^{\mu\nu}
    \left[
    \frac{1}{2}\,\partial^{\alpha}S\,\partial_{\alpha}S - V(S)
    \right],
    \label{eq:T0DM}
    \\
    T^{\mu\nu}_\Psi &=&
    \frac{i}{8}\left[
    \bar{\Psi}\gamma^{\mu}\overset{\leftrightarrow}{\partial^{\nu}}\Psi
    + \bar{\Psi}\gamma^{\nu}\overset{\leftrightarrow}{\partial^{\mu}}\Psi
    \right]
    -\nonumber\\&& \eta^{\mu\nu}\!\left[
    \frac{i}{4}\bar{\Psi}\gamma^{\alpha}\overset{\leftrightarrow}{\partial_{\alpha}}\Psi
    - \frac{m_\Psi}{2}\,\bar{\Psi^c}\,\Psi
    \right],
    \label{eq:T12DM}
    \\
    T^{\mu\nu}_X &=&
    -\frac{1}{2}\!\left[
    X_{\alpha}{}^{\mu}\,X^{\alpha\nu}
    + X_{\alpha}{}^{\nu}\,X^{\alpha\mu}
    - \frac{1}{2}\,\eta^{\mu\nu}\,X^{\alpha\beta}X_{\alpha\beta}
   \right.\nonumber\\&& \left.+ \eta^{\mu\nu}\,m_X^2\,X_{\alpha}X^{\alpha}
    - 2\,m_X^2\,X^{\mu}X^{\nu}
    \right],
    \label{eq:T1DM}
\end{eqnarray}
each of which contributes to \(T_{\rm DM}^{\mu\nu}\), one at a time.
\section{Frozen-in dark matter in warped extra dimensions}\label{sec:freezein}
\begin{figure}
    \centering
    \includegraphics[scale=0.086]{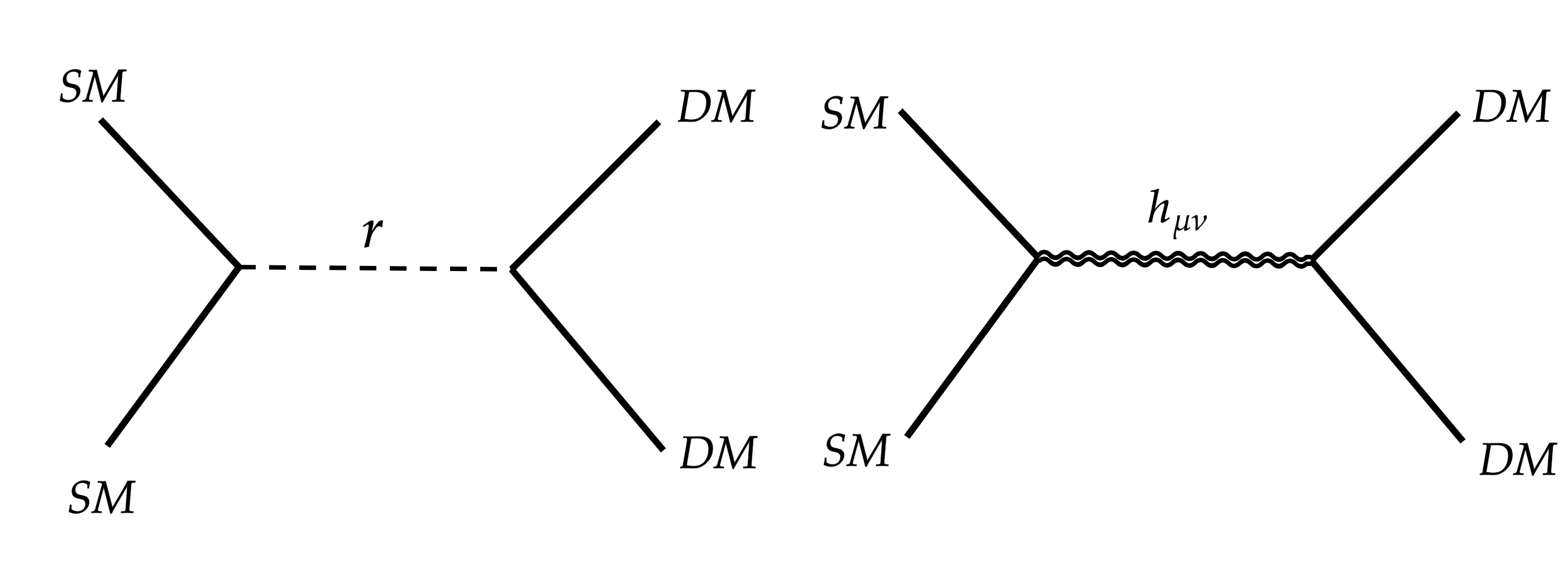}
    \caption{Freeze-in production of DM from scattering of the bath particles via radion portal (left) or graviton portal (right). The DM can be spin-0, spin-1/2 or spin-1 field. Similar processes also produce RHNs that participate in baryogenesis via leptogenesis.}
    \label{fig:feyn}
\end{figure}

The DM candidate under consideration is a feebly interacting massive particle, produced through the freeze-in process. Consequently, it never attains thermal equilibrium with the SM bath, and its abundance remains consistently below the equilibrium value throughout cosmic history. The central premise of freeze-in is the assumption of an initially empty dark sector in the early Universe. As time progresses, DM is gradually populated via interactions between the visible sector and the dark sector, enabled by feeble DM–SM couplings. In our analysis, we focus on DM production through scattering processes among SM particles, mediated by the radion and gravitons. We investigate three possible DM spin assignments: spin-0 scalar, spin-1/2 (Majorana) fermion, and spin-1 massive vector boson, considering each case independently. In the entire analysis we shall focus on direct freeze-in from the thermal bath, mediated by radion and graviton in the $s$-channel, as shown in Fig.~\ref{fig:feyn}.
\subsection{Dark matter genesis}
In order to follow the evolution of the DM number density $n_{\rm DM}$ with cosmic time, we begin with the Boltzmann equation (BEQ)\footnote{Including both inverse decays and scattering via KK gravitons leads to double counting of the on-shell contribution, since resonant $2\to2$ scattering through a KK mode is equivalent to its production via inverse decay followed by its subsequent decay in the narrow-width limit. In this limit, the Breit--Wigner propagator reduces to a delta function, and the total rate factorizes as $\Gamma_{\mathscr{G}\to{\rm SM}}\times {\rm BR}(\mathscr{G}\to{\rm DM})$. Therefore, one may consistently include either the inverse-decay channel or the full resonant scattering term, but not both, as pointed out in~\cite{deGiorgi:2022yha}.},  
\begin{align}
\dot n_{\rm DM}+3\mathcal{H}\,n_{\rm DM}=\gamma_{2\to2}\,,    
\end{align}
where, for a $2\!\to\!2$ scattering process, the reaction density takes the form~\cite{Duch:2017khv},
\begin{eqnarray}\label{eq:gamma22}
\gamma_{2\to2}&=&\frac{T\,g_a g_b}{32\pi^4}
\int_{\text{max}\left[\left(m_a+m_b\right)^2,\,4\mdm^2\right]}^\infty \!\! ds~\sigma_{a,b\to\text{DM},\text{DM}}(s)
\nonumber\\&&\times\frac{\bigl[(s-m_a^2-m_b^2)^2-4m_a^2 m_b^2\bigr]}{\sqrt{s}}\,
K_1\!\left(\frac{\sqrt{s}}{T}\right)\,,
\end{eqnarray}  
with $a,b$ denoting the incoming states, and $g_{a,b}$ their respective internal degrees of freedom. In a standard radiation-dominated epoch, where the entropy per comoving volume is conserved, it is convenient to recast the above equation in terms of the DM yield $Y_{\rm DM}\equiv n_{\rm DM}/s$,  
\begin{equation}\label{eq:beq}
x\,\mathcal{H}\,\mathfrak{s}\,\frac{dY_{\rm DM}}{dx} =\gamma_{2\to2}(T)\,,
\end{equation}
where $x \equiv m_{\rm DM}/T$ is a dimensionless variable with $T$ the bath temperature and 
\begin{align}
& \gamma_{2\to2}=\sum_{i\in r,\,\mathscr{G}}\gamma^i_{2\to2}\,, 
\end{align}
is the total reaction density for all graviton and radion mediated processes. For a radiation-dominated Universe, the entropy density $\mathfrak{s}(T)$ and the Hubble rate $H(T)$ are given by  
\begin{align}
&\mathfrak{s}(T)=\frac{2\pi^2}{45}\,\gss(T)\,T^3, \qquad 
\mathcal{H}(T)=\frac{\pi}{3}\,\sqrt{\frac{\gs(T)}{10}}\,\frac{T^2}{M_P}\,,
\end{align}
where $\gss(T)$ and $\gs(T)$ denote the effective relativistic degrees of freedom associated with entropy and energy density, respectively and $M_P$ is the reduced Planck mass.  The DM yield at temperature $T$ can then be obtained by integrating Eq.~\eqref{eq:beq},  
\begin{align}\label{eq:dm-yield}
Y_{\rm DM}(T)=-M_P \int_{\Trh}^T \mathcal{C}(T)\,\frac{\gamma_{2\to2}(T)}{T^6}\,,
\end{align}
with $\mathcal{C}(T)=\left[\left(2\pi^2/45\right)\gss(T)\sqrt{\pi^2 \gs(T)/90}\right]^{-1}$. Here, $\Trh$ is the reheating temperature defined under the sudden inflaton-decay approximation, as the maximum temperature of the thermal bath. In deriving Eq.~\eqref{eq:dm-yield}, a vanishing initial DM abundance at $T=\Trh$ has been assumed, as appropriate for freeze-in scenarios. To account for the observed relic abundance, the present-day DM yield must satisfy  
\begin{equation}\label{eq:obsyield}
Y_0\,m_{\rm DM} = \Omega h^2 \,\frac{1}{s_0}\,\frac{\rho_c}{h^2}\simeq  4.3\times 10^{-10}\,{\rm GeV}\,,
\end{equation}
where $Y_0 \equiv Y_{\rm DM}(T_0)$ is DM yield at present epoch, $\rho_c \simeq 1.05 \times 10^{-5} h^2\,{\rm GeV/cm}^3$ is the critical energy density, $s_0\simeq 2.69 \times 10^3\,{\rm cm}^{-3}$ the present entropy density~\cite{ParticleDataGroup:2022pth} and $\Omega h^2 \simeq 0.12$ the measured DM relic density~\cite{Planck:2018vyg}. 

To analytically understand the aspects of freeze-in, we work in the limit where the initial and final states have negligible masses with respect to the center of mass energy $\sqrt{s}$. This is a legitimate approximation for early times when $T\gg m$, typically above the EW symmetry breaking where all SM fields are massless. For the detailed analysis, we however solve the BEQ full numerically taking all masses and decay rates into account. In the massless limit, the total DM production cross-section via graviton and radion portal simplifies to,
\begin{eqnarray}
\sigma(s)_{\rm total}&\simeq& \frac{s^3}{512\,\pi\,\Lambda_r^4}\,\frac{1}{(s-m_r^2)^2+\Gamma_r^2\,m_r^2}+\nonumber\\&&\frac{107\,s^3}{51840\pi\,\Lambda_r^4}\,\left|\sum_n\frac{1}{(s-m_{G_n}^2)+i\,\Gamma_n\,m_{G_n}}\right|^2\,,  
\end{eqnarray}
considering scalar DM scenario and using the cross-sections listed in Appendix.~\ref{sec:rad-med-cs} and \ref{sec:grav-med-cs}. Note that, in the above expression we have considered contribution from the radion and only from higher KK graviton modes. In this limit, the reaction rate density in Eq.~\eqref{eq:gamma22} also simplifies to
\begin{align}
&\gamma_{2\to2}\simeq g_a\,g_b\frac{T}{32\,\pi^4}\,\int_{4\,\mdm^2}^\infty ds\,s^{3/2}\,\sigma(s)\,K_1\left(\frac{\sqrt{s}}{T}\right)\,.   
\end{align}
It is worth mentioning that while for radion-mediation the only available DM production channel above EWSB is $hh\to\text{DM}\,\text{DM}$, for graviton meditation we have all possible (massless) SM initial states. Near the resonance, one can further apply the narrow width approximation (NWA) that results in,
\begin{align}
&  \frac{1}{(s-m_{\rm med}^2)^2+\Gamma_{\rm med}^2\,m_{\rm med}^2}\to\frac{\pi}{m_{\rm med}\,\Gamma_{\rm med}}\,\delta\left(s-m_{\rm med}^2\right)\,,   
\end{align}
where `med' implies both radion and graviton mediator. Away from the resonance the propagator becomes $1/(s-m_{\rm med}^2)^2$, ignoring the mediator decay rate. With the approximations and assumptions furnished, we obtain an approximate analytical expression for the DM yield via 2-to-2 scattering as,
\begin{widetext}
\begin{align}\label{eq:gamma-analyt}
&\gamma_{2\to2}\simeq\frac{1}{\Lambda_r^4}
\begin{dcases}
\frac{5\,T^{12}}{m_r^4\,\pi^5} & T \ll m_r/2\,,
\\[4pt]
\frac{m_r^8\,T}{589824\,\pi^4\,\Gamma_r}\times K_1\left(m_r/T\right)     & T \simeq m_r/2\,,
\\[4pt]
\frac{T^8}{768\,\pi^5}     & m_r/2 \ll T \ll m_{G_1}/2\,,
\\[10pt]
\frac{899\,m_{G_1}^8\,T}{414720\,\pi^4\,\Gamma_1}\times K_1\left(m_{G_1}/T\right)\,   & T \simeq m_{G_1}/2\,,
\\[4pt]
\frac{925\,x_1}{36864\,\pi^3}\,\frac{m_{G_1}^2\,T^7}{\Gamma_1}  & T \gg m_{G_1}/2\,,
\\[4pt]
\end{dcases}
\end{align}
\end{widetext}
where $K_1(z)$ is the modified Bessel function of the first kind. In the last line we have  approximated the KK tower as a continuum and replace $\sum_n\to\int dm_n/\Delta m$ in the propagator~\cite{Han:1998sg,deGiorgi:2022yha}, where $\Delta m\simeq m_{G_1}/x_1$, with $x_1\simeq 3.8$ [cf. Eq.~\eqref{eq:mn}]. Utilizing Eq.~\eqref{eq:gamma-analyt}, it is then possible to solve Eq.~\eqref{eq:beq} analytically and compute the final DM yield via radion mediation as,

\begin{widetext}
\begin{align}\label{eq:yld-analytic}
& Y_0\simeq\frac{M_P\,\mathcal{C}}{\Lambda_r^4} 
\begin{dcases}
\frac{5\,\Trh^7}{7\,\pi^5\,m_r^4} & \Trh\ll m_r/2\,,
\\[10pt]
\frac{e^{-m_r/\Trh}\,m_r^{9/2}}{2359296\,\pi^4\,\Trh^{5/2}\,\Gamma_r}\,\left(4\,m_r^2+10\,m_r\,\Trh+15\,\Trh^2\right) & \Trh\simeq m_r/2\,,
\\[10pt]
\frac{\Trh^3}{2304\,\pi^5} & m_r/2\ll\Trh\ll m_{G_1}/2\,,
\\[10pt]
\frac{899\,e^{-m_{G_1}/\Trh}\,m_{G_1}^{9/2}}{1658880\,\pi^4\,\Trh^{5/2}\,\Gamma_1}\,\left(4\,m_{G_1}^2+10\,m_{G_1}\,\Trh+15\,\Trh^2\right) & \Trh\simeq m_{G_1}/2\,,
\\[10pt]
\frac{925}{73728\,\pi^3}\,\frac{m_{G_1}^2\,\Trh^2}{\Gamma_1} & \Trh\gg m_{G_1}/2\,,
\end{dcases}
\end{align}
\end{widetext}
where we have used the asymptotic approximation: $K_\nu(x)\sim x^{-1/2}\,e^{-x}$, which holds for $x\to\infty$ or, in other words, $T\to0$ for any $\nu$. Note that, typically for $T\gtrless m_{\rm med}/2$, the DM yield has a strong dependence on $\Trh$, i.e., bulk of the DM is produced at $T\sim\Trh$, a quintessential feature of UV freeze-in. Note that, in deriving Eq.~\eqref{eq:yld-analytic} we have considered $\mathcal{C}(T)=\mathcal{C}$, i.e., the relativistic DoFs do not evolve with temperature. We emphasize that although Eq.~\eqref{eq:yld-analytic} has been calculated for scalar DM, however, the same expressions also hold true for spin-1 or spin-1/2 DM, with only modification in the numerical pre-factors. It is essential for freeze-in that the DM production rate remains below the Hubble rate during production. Since in the present scenario, bulk of the DM is produced around $T\simeq \Trh$, hence we need to ensure 
\begin{align}
& \frac{\gamma_{2\to2}}{n_{\rm eq}^{\rm DM}}\Bigg|_{T=\Trh}<\mathcal{H}(\Trh)\,,    
\end{align}
where $n_{\rm eq}^{\rm DM}(T)=\frac{g\,T}{2\pi^2}\,\mdm^2\,K_1\left(\frac{\mdm}{T}\right)$ is the equilibrium number density, with $g$ being the internal DoF of the concerned species. With this, one can obtain an upper bound on the reheating temperature, below which out of equilibrium freeze-in production can be ensured,
\begin{widetext}
\begin{align}\label{eq:trh}
&\Trh < 
\begin{dcases}
\left[\frac{(\pi\Lambda_r\,m_r)^4}{M_P}\,\frac{7\,\sqrt{\gs}}{15}\right]^{1/7} & \Trh\ll m_r/2\,,
\\[10pt]
-\frac{2\,m_r}{7}\,\left(\mathcal{W}\left[-\frac{2}{7}\,\left(\frac{\Gamma_r^2\,\Lambda_r^8}{M_P^2\,m_r^8\,\mathcal{A}^2}\right)^{1/7}\right]\right)^{-1} & \Trh\simeq m_r/2\,,
\\[10pt]
\left[16\,\sqrt{\gs}\,\frac{(\pi\Lambda_r)^4}{M_P}\right]^{1/3} & m_r/2\ll\Trh\ll m_{G_1}/2\,,
\\[10pt]
-\frac{2\,m_{G_1}}{7}\,\left(\mathcal{W}\left[-\frac{2}{7}\,\left(\frac{\Gamma_1^2\,\Lambda_r^8}{M_P^2\,m_{G_1}^8\,\mathcal{B}^2}\right)^{1/7}\right]\right)^{-1} & \Trh\simeq m_{G_1}/2\,,
\\[10pt]
\sqrt{\frac{24576\,\pi^2\,\gs^{1/2}}{925\,x_1}}\,\sqrt{\frac{\Gamma_1}{M_P}}\,\frac{\Lambda_r^2}{m_{G_1}} & \Trh\gg m_{G_1}/2\,,
\end{dcases}
\end{align}
\end{widetext}
where $\mathcal{A}=1.6\times 10^{-7}/\sqrt{\gs}$, $\mathcal{B}\simeq 2.1\times 10^{-4}/\sqrt{\gs}$ and $\mathcal{W}$ is the Lambert W-function. This is rather a conservative bound since we have considered an equilibrium number density for the DM. We solve Eq.~\eqref{eq:beq} numerically by taking into account all possible channels above and below the EWSB temperature $T=T_{\rm EW}\simeq 160$ GeV, properly tracking the relativistic DoFs via $\gs(\gss)(T)$. As stated before, all SM states are considered to be absolutely massless for $T>T_{\rm EW}$, while for $T<T_{\rm EW}$ they become massive. Before moving on let us clarify that following are the independent parameters in the present framework:
\begin{align}
\{\mdm,\,m_r,\,\Lambda_r,\,\Trh\}\,,
\end{align}
whilst the graviton mass is fixed by Eq.~\eqref{eq:mn}.
\begin{figure*}[tbh]
    \centering        \includegraphics[scale=0.42]{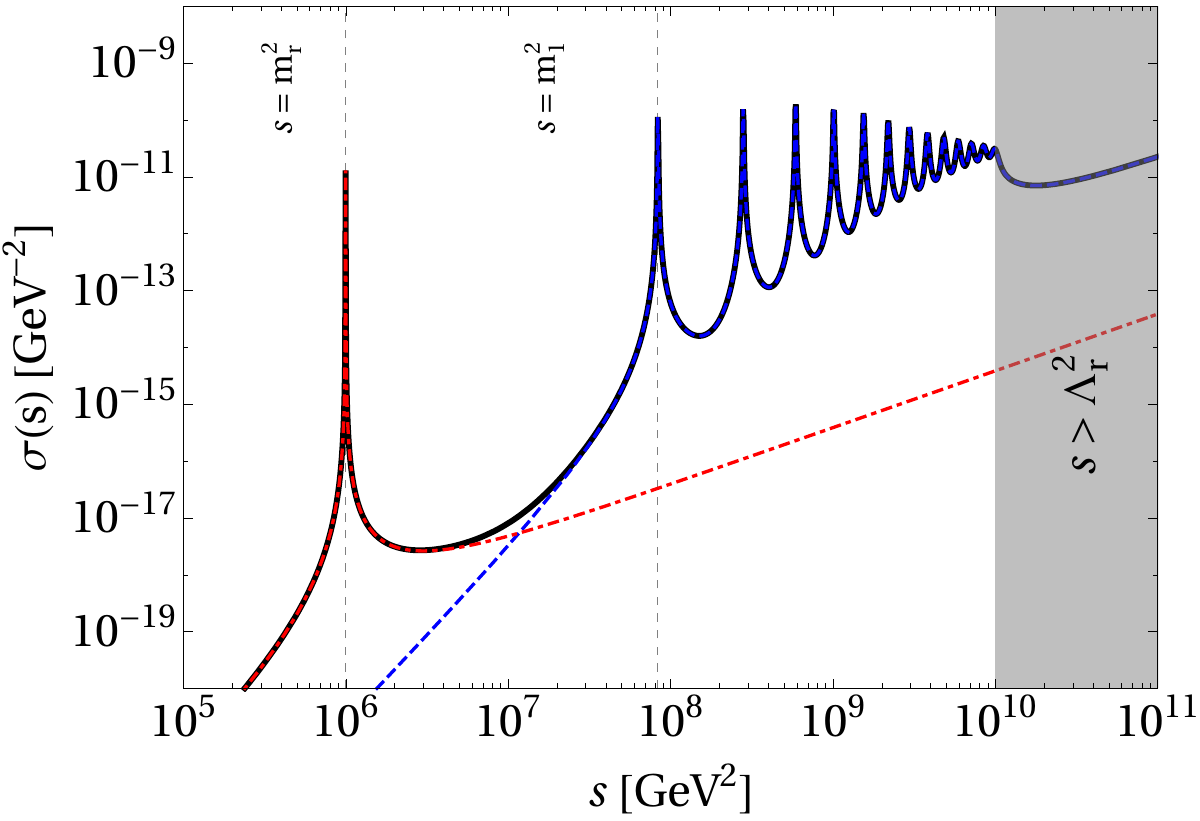}~\includegraphics[scale=0.42]{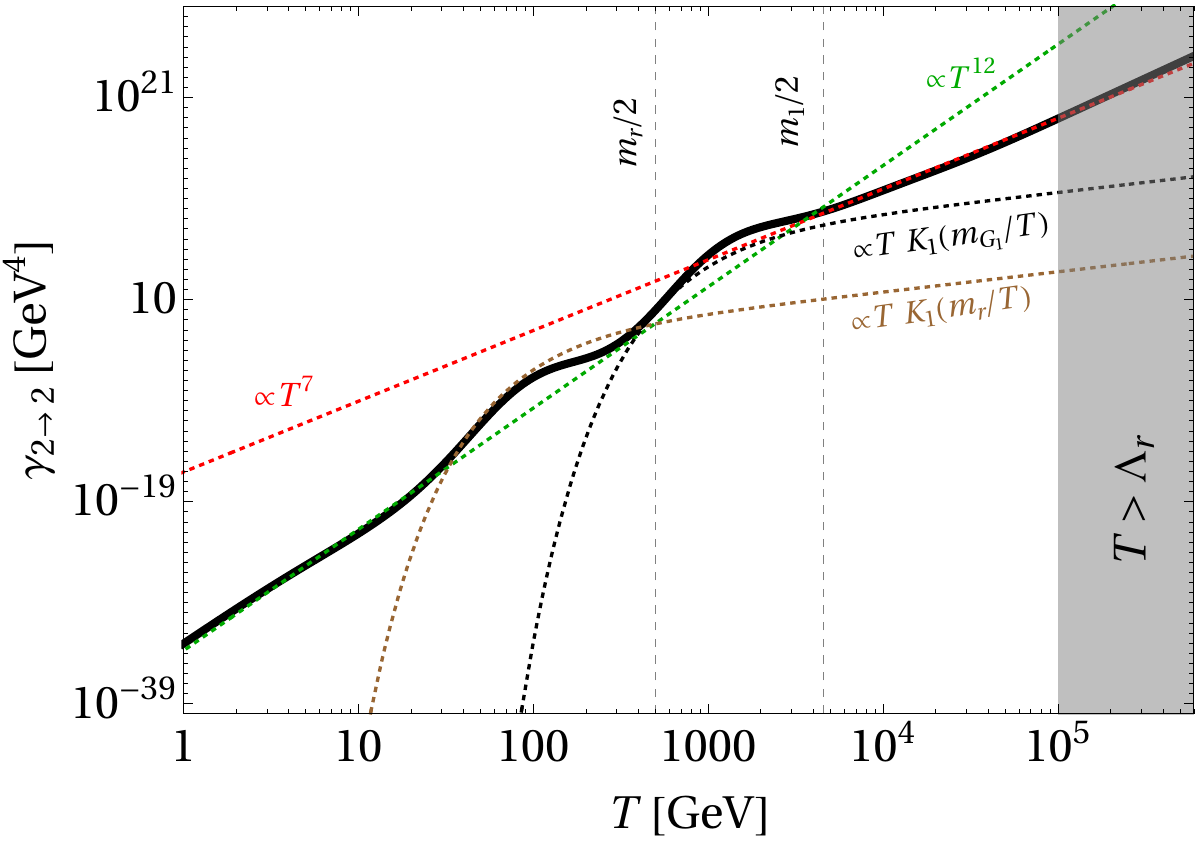}
    \caption{Left: DM production cross-section as function of $s$, taking into account both radion and graviton exchange. Right: Reaction density as a function of temperature $T$. We fix $\Lambda_r=100$ TeV, $\mdm=1$ GeV, $m_r=1$ TeV, $kr_c=11$ (corresponding graviton mass can be derived from Tab.~\ref{tab:mn}) and consider scalar DM. The shaded regions violate the effective theory condition.}
    \label{fig:reacdensity}
\end{figure*}

The left panel of Fig.~\ref{fig:reacdensity} displays the DM production cross section as a function of the center-of-mass energy $\sqrt{s}$, including both radion- and graviton-mediated channels for spin-0 DM. The first resonance originates from radion exchange, while the subsequent peaks correspond to KK graviton modes. The characteristic $\sim s^3$ growth for graviton mediation is evident. Our benchmark parameters are chosen such that the hierarchy problem is consistently addressed. The radion peak is sharper than the graviton ones owing to its narrower width, though the distinct graviton resonances are still discernible. At energies $\sqrt{s} \gg m_1$, these resonant features fade, and the cross section transitions into a continuum primarily governed by graviton exchange. The right panel shows the temperature dependence of the $2\to 2$ reaction density. Four regimes are visible: (i) for $T\ll m_r/2$, mediators are too heavy to participate, yielding $\gamma\propto T^{12}$ (green dotted); (ii) near $T\simeq m_r/2$, radion exchange dominates, producing $\gamma\propto T K_1(m_r/T)$ (brown dotted), corresponding to the first bump; (iii) at $T\simeq m_{G_1}/2$, the first KK graviton mode takes over, with $\gamma\propto T K_1(m_{G_1}/T)$, giving the second bump (black dotted); and (iv) for $T\gg m_{G_1}/2$, the mediator mass becomes negligible, leading to $\gamma\propto T^7$ (red dotted). With $m_r=1$~TeV and $m_{G_1}\simeq 9$~TeV (fixed by $kr_c=11$), the two bumps lie close together, so the intermediate $T^8$-scaling is not distinctly visible. The maximum contribution to the reaction density due to gravitons comes from the first KK-mode. Altogether, the analytic behavior of Eq.~\eqref{eq:gamma-analyt} closely reproduces the numerical scaling of the reaction density. The shaded region in both plots is beyond the EFT approach, as the center-of-mass energy of the process (in the left panel) or the temperature of the radiation bath (in the right panel) is larger than the effective scale of the theory.

\begin{figure*}[tbh]
    \centering
    \includegraphics[scale=0.42]{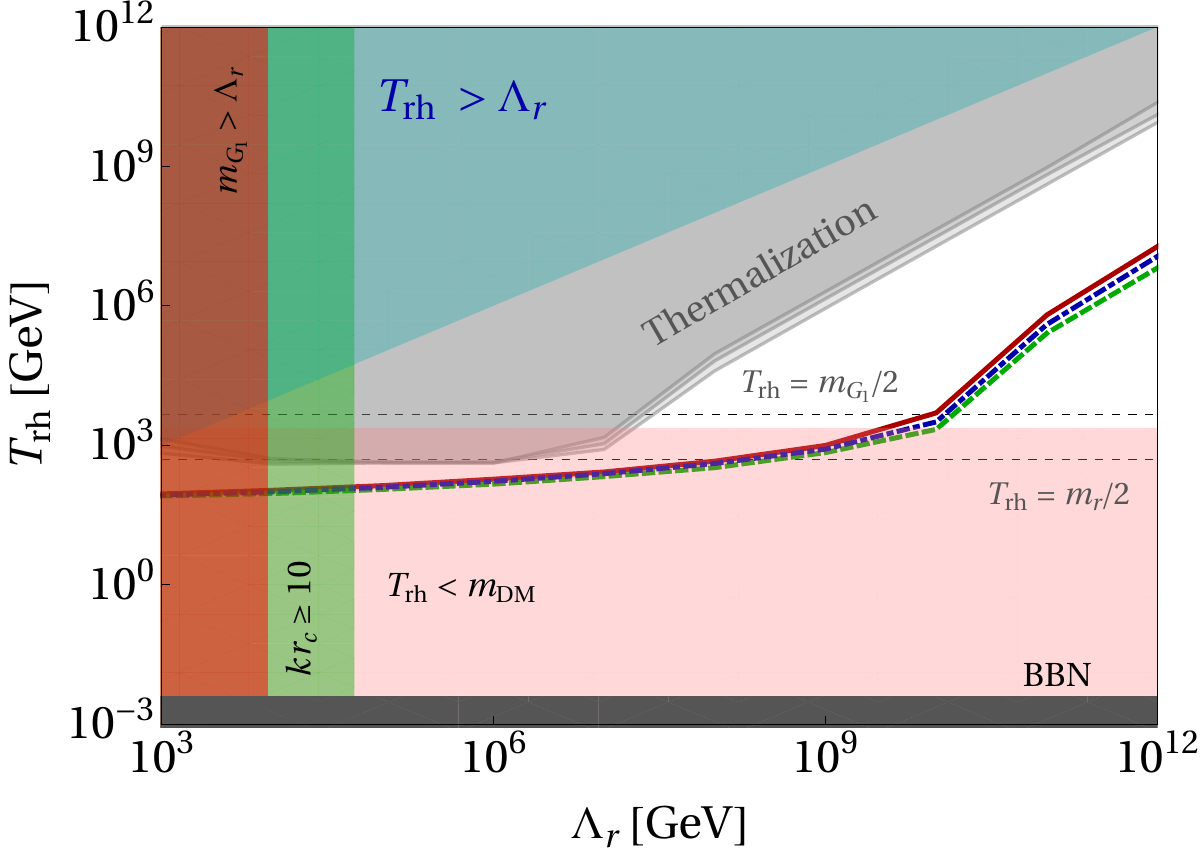}
    \includegraphics[scale=0.42]{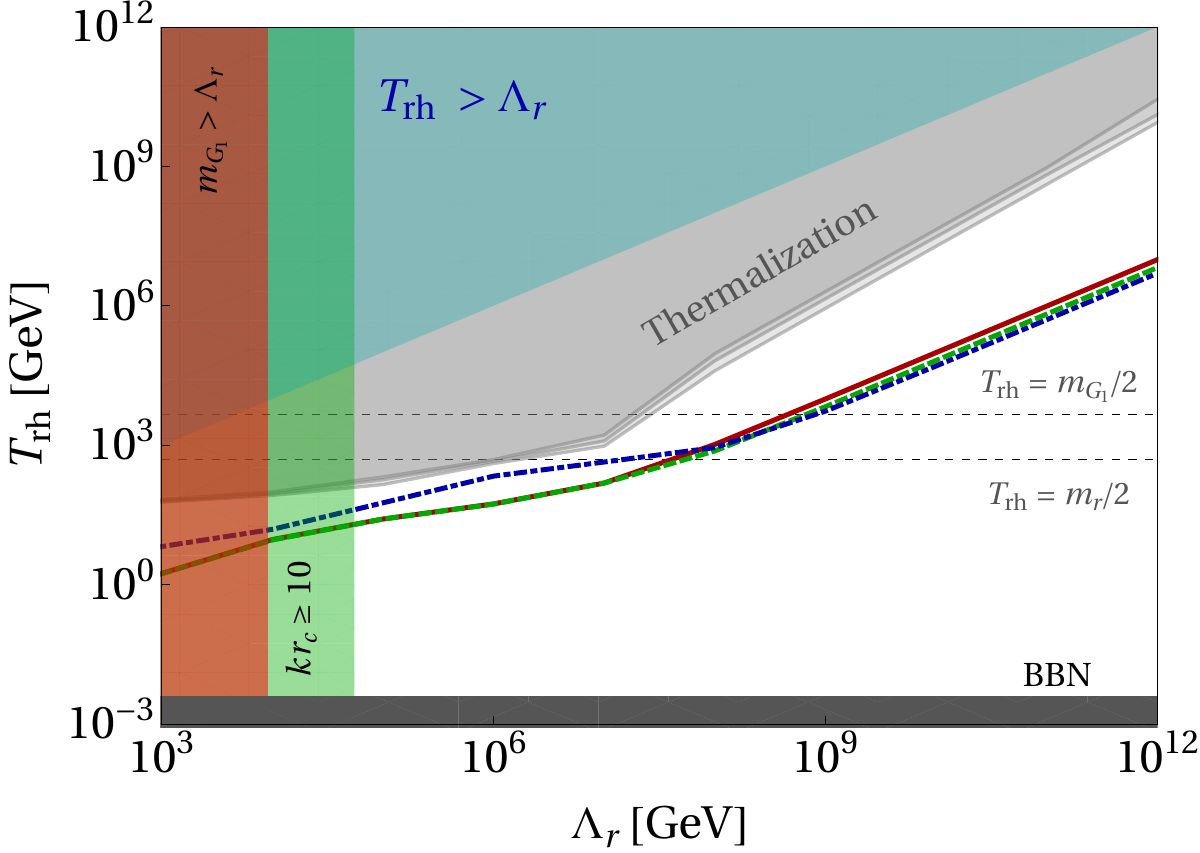}
    \caption{Contours of right relic abundance for a DM of mass $\mdm\simeq2300$ GeV, with corresponding $kr_c=11$, shown in the left panel, considering different DM spins. The red solid, green dashed and blue dot-dashed contours correspond to spin-0, spin-1 and spin-1/2 DM, respectively. The right panel shows the same, but for a DM of mass 1 MeV. In all cases we have fixed $m_r=1$ TeV and follow Tab.~\ref{tab:mn} for the graviton mass. We show regions excluded from DM thermalization condition (in lighter gray), BBN bound on the reheating temperature (in darker gray), instantaneous reheating condition (in pink, in the top panel) and violation of effective theory (in cyan \& in darker red). Within the green shaded region, the hierarchy problem can be addressed.}
    \label{fig:1}
\end{figure*}

Contours corresponding to the observed DM abundance for fixed DM masses are shown in Fig.~\ref{fig:1}, in the bi-dimensional plane of $[\Trh-\Lambda_r]$. For each case, we display all three possible DM spins. In the upper panel, we set $k r_c = 11$ and fix the DM mass by taking $m_S^0 = M_P$. The implications of varying the DM mass will be discussed shortly. Irrespective of the DM spin, the parameter spaces exhibit a broadly similar pattern. This behavior can be understood from the approximate analytical expression in Eq.~\eqref{eq:yld-analytic}, which shows that the DM relic abundance scales as $\Omega_{\rm DM} h^2 \sim m_{\rm DM}\, \Trh^{\xi} / \Lambda_r^4$,
where $\xi > 0$ depends on the specific temperature regime under consideration. Consequently, for a fixed DM mass, an increase in $\Lambda_r$ leads to an underabundance, which can be compensated by a larger reheating temperature $\Trh$. However, excessively large $\Trh$ values may lead to thermalization of the SM--SM $\leftrightarrow$ DM--DM processes (shown in dark gray), thereby violating the freeze-in assumption. Since we assume instantaneous reheating, the SM plasma cannot produce DM heavier than the reheating temperature. This excludes the pink-shaded regions in the upper panel. It is however worth noting that DM production
can proceed even if $\mdm>\Trh$, which relies on the Boltzmann tail of the SM particle distribution that enables the generation of the DM relic at low temperature~\cite{Lee:2024wes,Cosme:2023xpa,Bernal:2025fcl}. For $\mdm=1$ MeV, shown in the lower panel, this constraint is absent because the BBN bound $T_{\rm BBN} \simeq 4~\text{MeV} > m_{\rm DM}$ is always satisfied. For heavier DM, one observes that as $\Lambda_r$ increases, the required $\Trh$ also rises, although more slowly than for lighter DM, as seen by comparing the upper and lower panels. This difference arises because for lighter DM, phase-space suppression is negligible, and the yield grows rapidly as $Y_{\rm DM} \propto \Trh^7$, following the first line of Eq.~\eqref{eq:yld-analytic}. As a result, for the same $\Trh$, lighter DM requires a larger $\Lambda_r$ to avoid overproduction, an effect mitigated by increasing the scale $\Lambda_r$. For the $\mdm=1$ MeV case, one can still take $k r_c = 11$ with $m_S^0 \simeq 10^{12}~\text{GeV}$. It is worth emphasizing that, for a given $\Lambda_r$ and $k r_c$, the reheating temperature cannot be arbitrarily large without invalidating the 4D effective theory. The corresponding exclusion region, where the effective description is expected to fail, is shown in cyan. Nevertheless, as evident from the figure, thermalization typically imposes a stronger constraint. The green-shaded region corresponds to $k r_c \leq 10$, which can be translated into $\Lambda_r \lesssim 5.5 \times 10^4~\text{GeV}$. Consequently, for heavier DM, the simultaneous realization of the correct relic abundance and resolution of the hierarchy problem is ruled out under the assumption of instantaneous reheating, that forbids $\Trh<\mdm$. In contrast, for lighter DM, as illustrated in the lower panel, a consistent solution addressing both the hierarchy problem and the observed DM abundance remains viable. As mentioned earlier, a modest variation in the bulk curvature $k$ (as well as $kr_c$) can lower the graviton mass scale to values accessible at colliders. However, since our focus is not on collider phenomenology, we choose the model parameters such that the corresponding masses remain well above the current collider bounds (see subsection~\ref{sec:constraints} for a discussion on constraints). It is worth noting that varying the mediator masses within this range does not qualitatively affect our results.
\begin{figure*}[tbh]
    \centering
    \includegraphics[scale=0.42]{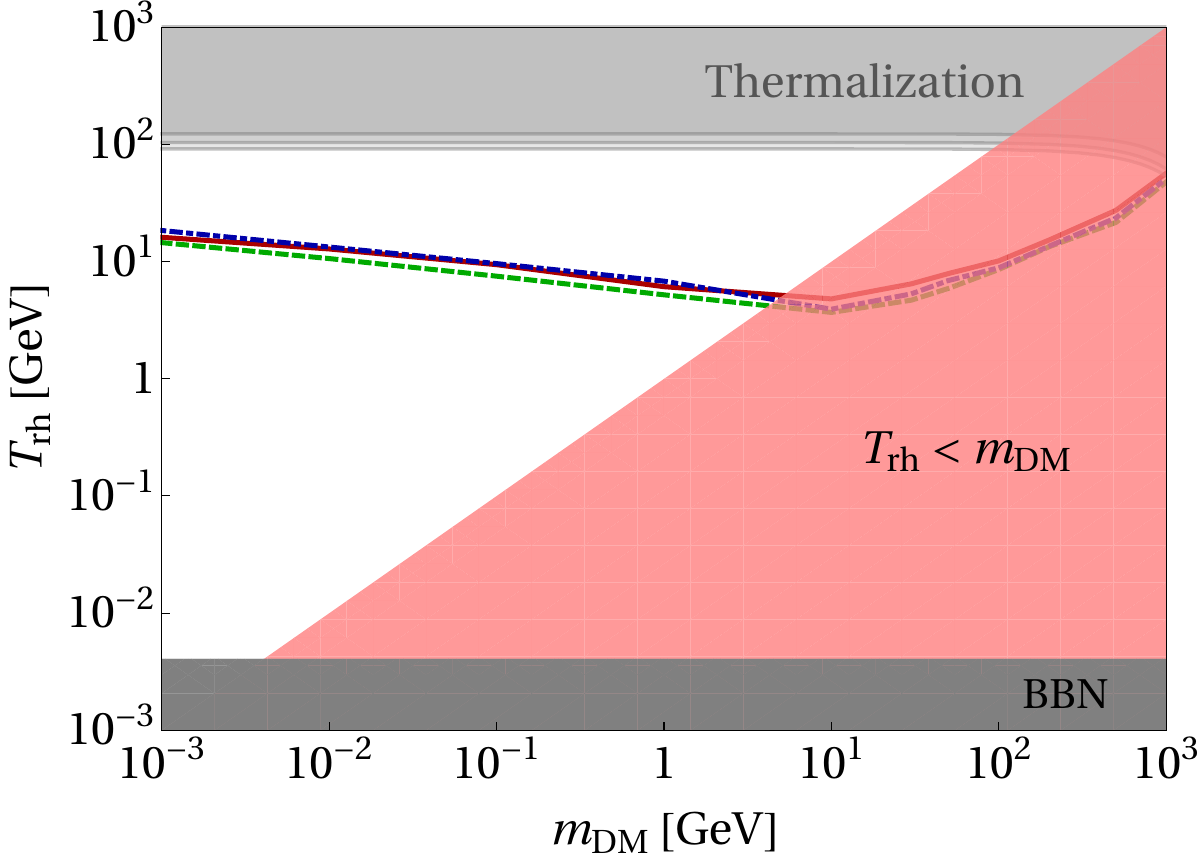}~\includegraphics[scale=0.34]{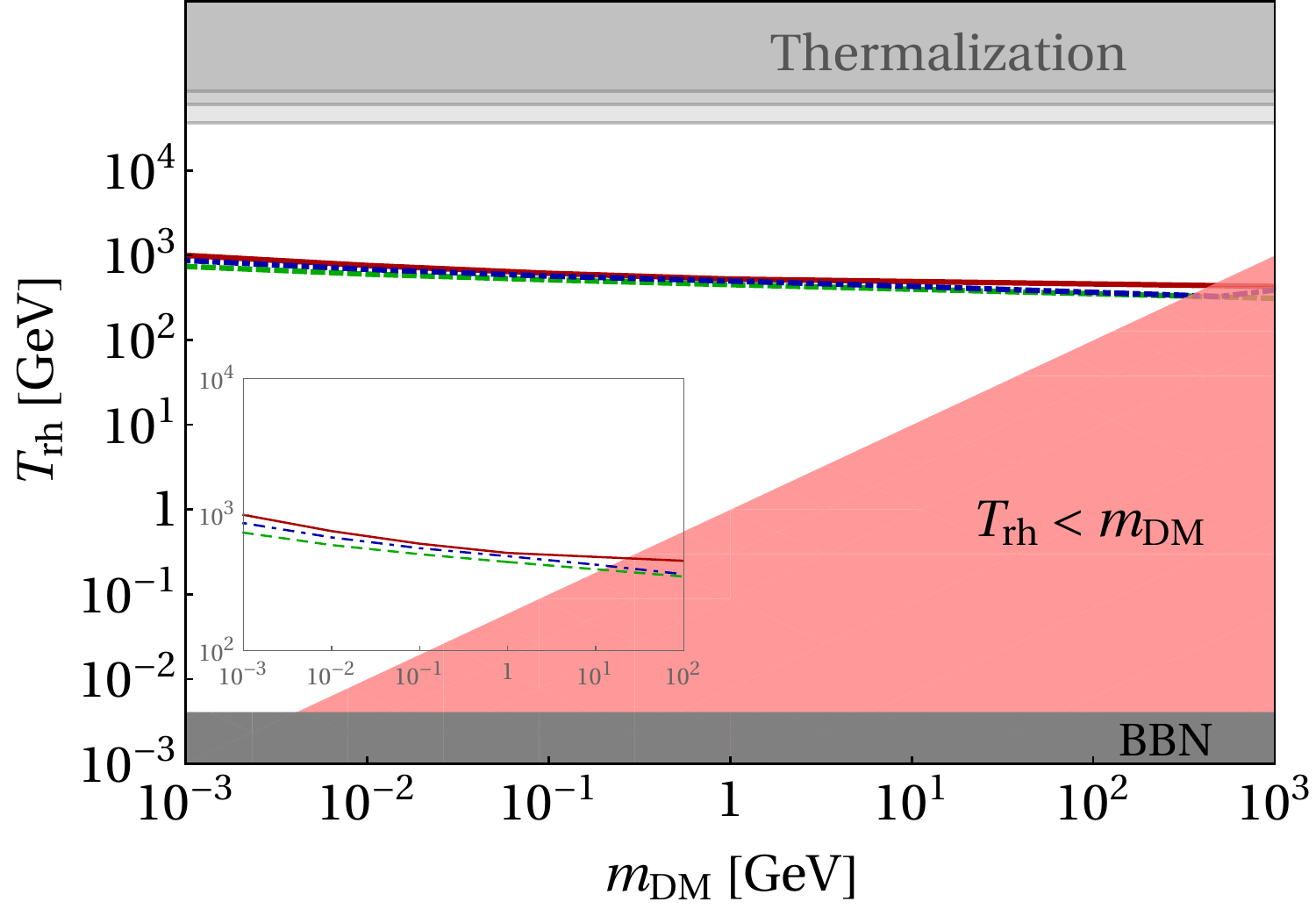}
    \caption{Contours of right DM abundance for different DM spins, considering $m_r=1$ TeV, $\Lambda_r=10$ TeV in the left panel and $\Lambda_r=10^5$ TeV in the right panel. The red solid, green dashed, and blue dot-dashed contours correspond to spin-0, spin-1, and spin-1/2 DM, respectively. Different shaded regions are excluded from the DM thermalization condition (in lighter gray), the BBN bound on the reheating temperature (in darker gray), and the instantaneous reheating condition (in lighter red). In the right panel inset, we have zoomed into the relic density contours corresponding to different spins to show them more clearly.}
    \label{fig:2}
\end{figure*}

We choose two benchmark $\Lambda_r$ values while fixing $m_r=1$ TeV, and obtain the allowed parameter space in $[\Trh-\mdm]$ plane, as shown in Fig.~\ref{fig:2}. Once again, part of the relic density allowed region is forbidden from the thermalization condition, typically for larger $\Trh$. As the DM mass keeps increasing, in order to satisfy the correct abundance, $\Trh$ decreases until the DM becomes heavy enough where it can no longer produced from the thermal bath. Note that, a larger $\Lambda_r$ requires higher $\Trh$ for the same DM mass, since in that case the DM production is suppressed and could only be overcome by considering a higher $\Trh$. As we will argue later, a larger $\Trh$ is compatible with the generation of baryon asymmetry. It is interesting to note that for $\Lambda_r=10$ TeV, one can still have $kr_c\sim\mathcal{O}(10)$, while keeping $10^{11}\lesssim m_S^0\lesssim 10^{17}$ GeV to obtain DM over MeV-TeV mass range on the brane [cf. Eq.~\eqref{eq:mdm1}], thereby providing a consistent resolution to both the hierarchy problem and the right DM abundance via freeze-in. Before concluding this section, it is useful to highlight an important aspect of the RS framework. As discussed earlier, the extra dimensional modulus in the RS model can be stabilized at a value close to the inverse Planck scale by introducing a massive scalar field in the bulk. In the original stabilization mechanism, however, the backreaction of this bulk scalar on the background geometry is neglected. This framework was later extended to a more general warped geometry setup by consistently including the backreaction of the stabilizing scalar field on the background metric (see, for example, Refs.~\cite{Csaki:2000zn,Grzadkowski:2003fx,Das:2015zxa,Chivukula:2024nzt}).  Following Ref.~\cite{Csaki:2000zn}, one finds that the condition $m_r/\Lambda_r \ll 1$ is naturally satisfied in the limit of small backreaction. In our analysis, the radion mass is fixed at $m_r = 1$ TeV throughout, while the values of $\Lambda_r$ within the viable parameter space of both Fig.~\ref{fig:1} and Fig.~\ref{fig:2} that yield the correct relic abundance always satisfy $m_r/\Lambda_r \ll 1$. Therefore, all benchmark points considered in this work consistently lie within the small backreaction regime.

\section*{Minimal gravitational freeze-in}
It is noteworthy that the DM is inevitably produced via exchange of the massless gravitons, due to the first term in Eq.~\eqref{eq:Lgraviton} which leads to irreducible DM-graviton interactions via {\it minimal} gravity. The interaction rate density for such a process reads~\cite{Garny:2015sjg,Tang:2017hvq,Garny:2017kha,Bernal:2018qlk,Barman:2021ugy}
\begin{equation}
\gamma(T)=k\,\frac{T^8}{M_P^4}\,,
\end{equation}
with $k\simeq 2.9\times 10^{-3}$ (spin-0), $k\simeq 1.7\times 10^{-2}$ (spin-1/2 Majorana) or  $k\simeq 7.3\times 10^{-2}$ (spin-1). In the sudden decay approximation for the inflaton, one can analytically obtain the DM asymptotic yield as~\cite{Barman:2021ugy},
\begin{align}
& Y_0\simeq\frac{45\, k}{2\pi^3\, \gss} \sqrt{\frac{10}{\gs}} \left(\frac{\Trh}{M_P}\right)^3\,,    
\end{align}
which leads to
\begin{equation}
\Trh\simeq\left(\frac{10^3\,\text{GeV}}{\mdm}\right)^{1/3}
\begin{dcases}
9.4\times10^{15}\,\text{GeV}\,, & \text{spin-0}\,,
\\[10pt]
5.2\times10^{15}\,\text{GeV}\,, & \text{spin-1/2}\,,
\\[10pt]
3.2\times10^{15}\,\text{GeV}\,, & \text{spin-1}\,,
\end{dcases}
\end{equation}
in order to satisfy the right DM abundance. Consequently, DM production via massless graviton mediation becomes important at a very high $\Trh$, because of the large hierarchy $\Lambda_r\ll M_P$. Thus, for $\Trh< 10^{15}$ GeV, the contribution from massless graviton mediation to the DM yield can be safely ignored. 
\subsection{Theoretical and observational constraints}
\label{sec:constraints}
\subsubsection{From collider}
From theoretical analyses, Ref.~\cite{Bae:2000pd} reported a radion mass range of 
$0.8~{\rm GeV} \lesssim m_r \lesssim 260~{\rm GeV}$ with 
$1.4\lesssim \Lambda_r \lesssim 1.5$ TeV. This parameter space corresponds to a radion lighter than the first KK graviton mode and is compatible with the warp factor required to reproduce the electroweak scale. Collider searches, however, impose far stronger bounds, as examined in~\cite{Bae:2000pk,Mahanta:2000ci,Cheung:2003ze,Cheung:2003fz,Cheung:2005pg,deSandes:2011zs,Ohno:2013sc,Cho:2013mva,Kubota:2014mma}. For instance, using $h\to ZZ$, $h\to W^+W^-$, and $h\to \gamma\gamma$ searches, Ref.~\cite{Cho:2013mva} demonstrated that at 
$m_r\simeq 200$ GeV the effective scale must satisfy $\Lambda_r\gtrsim 5$ TeV, while for $m_r=1$ TeV the lower limit reduces to about $2$ TeV. As emphasized in~\cite{Giudice:2000av,Csaki:2000zn,Dominici:2002jv}, the radion can also mix with the SM Higgs boson, with important phenomenological implications. Without such mixing, the only means of suppressing radion couplings is to raise its scale, which pushes LHC bounds on $\Lambda_r$ to multi-TeV values~\cite{deSandes:2011zs,Frank:2011kz,Barger:2011qn}. Including Higgs--radion mixing, however, enriches the phenomenology, enabling consistency with Higgs data and even evading exclusion limits without requiring a large new-physics scale~\cite{deSalas:2015glj,Desai:2013pga,Chakraborty:2017lxp,Sachdeva:2019hvk}. The mass bound on the lightest graviton excitation in this framework has already surpassed $2$ TeV at $\sqrt{s}=8$ TeV~\cite{CMS:2013egk,ATLAS:2014pcp}. The second theoretical aspect involves ensuring the consistency of the EFT framework. In the RS model, energies above $\Lambda_r$ lead to strong coupling among KK gravitons, signaling the breakdown of the five-dimensional description. We thus require {\it at least} the first KK graviton to remain lighter than the cutoff, $m_{G_1} < \Lambda_r$~\cite{Folgado:2019sgz}, shown by the red shaded area in Fig.~\ref{fig:1}. The ATLAS data corresponding to high-mass diphoton final states corresponding to an integrated luminosity of 36.7 fb$^{-1}$ at a center-of-mass energy of $\sqrt{s}=13$ TeV, has excluded graviton mass below 4.1 TeV for $k/M_P=0.1$~\cite{ATLAS:2017ayi}. A more recent analysis from CMS for the same signal event at $\sqrt{s}=13$ TeV for an integrated luminosity of 138 fb$^{-1}$ excludes masses below 2.2 to 5.6 TeV at 95\% confidence level, for $0.01<k/M_P<0.2$~\cite{CMS:2024nht}. The DM can, in principle, scatter off the SM nucleons via 
the $t$-channel exchange of the radion or graviton, leading to the possibilities of direct detection. However, the resulting constraints on $\Lambda_r$, from experiments such as LUX-ZEPLIN~\cite{LZ:2024zvo} and DARWIN~\cite{DARWIN:2016hyl},
are many orders of magnitude weaker than those derived from freeze-in.    
\begin{figure*}[tbh]
    \centering    \includegraphics[scale=0.42]{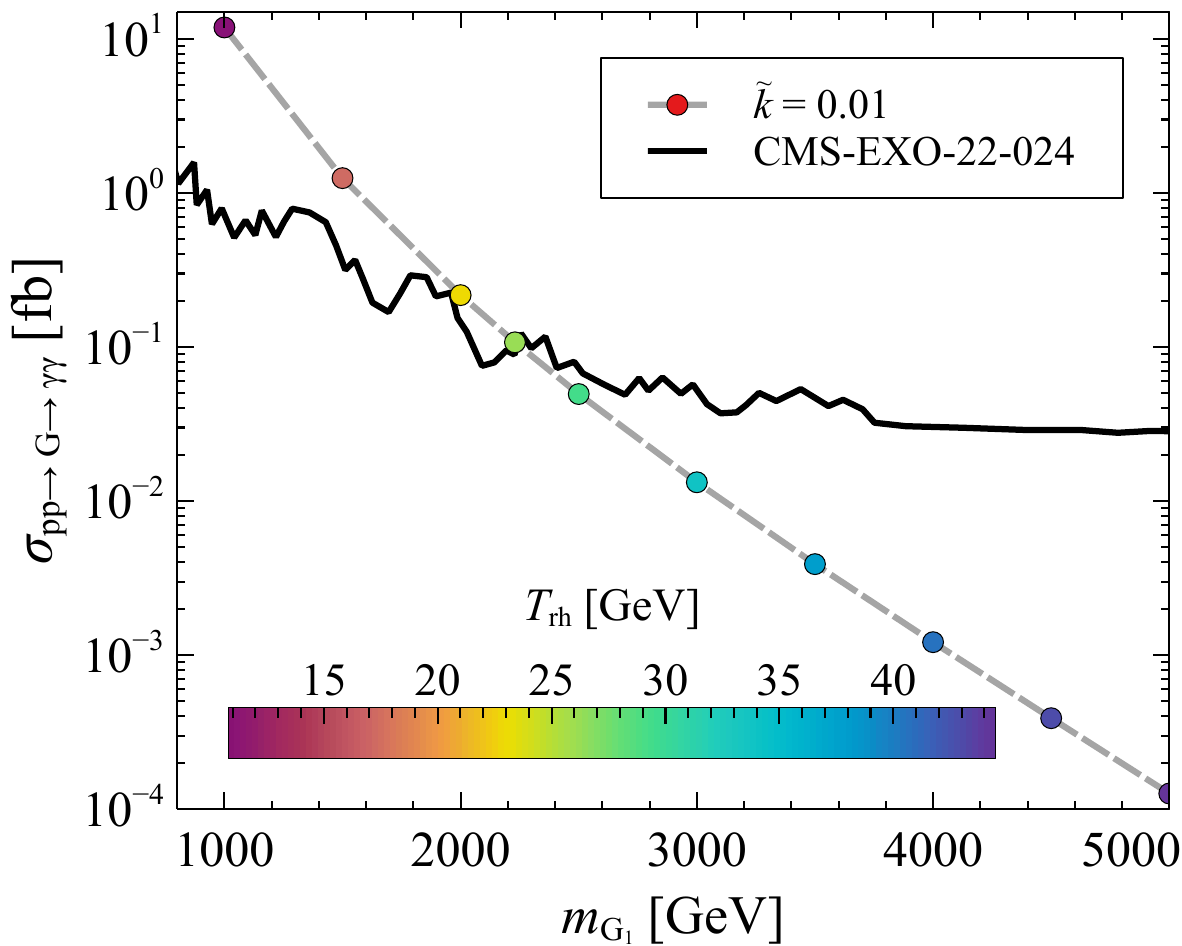}~   \includegraphics[scale=0.42]{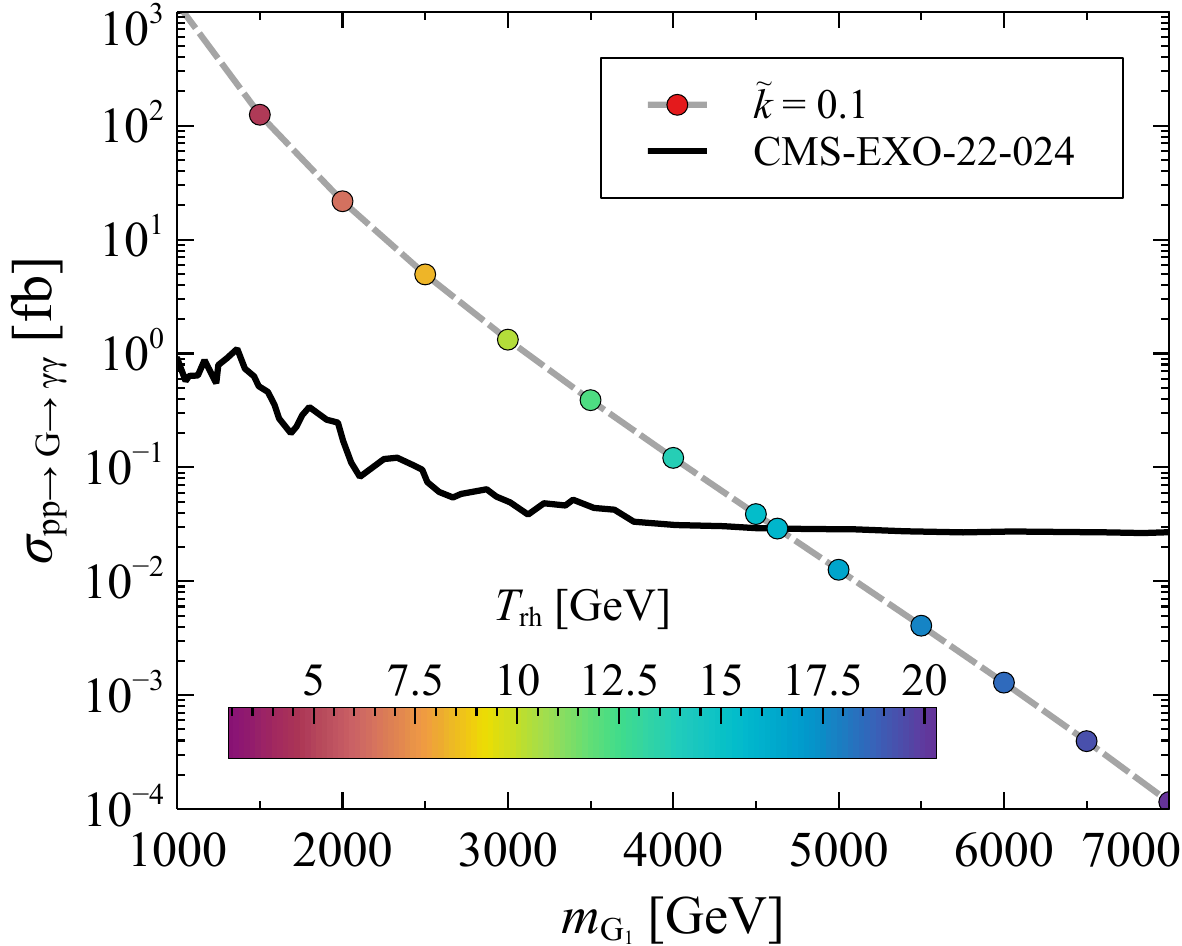}
    \caption{Production cross-section in high-mass diphoton $(pp\to\gamma\gamma)$ events from proton-proton collisions is shown (gray dashed line), as a function of the lightest KK-graviton mass. The black solid curve corresponds to result obtained from CMS~\cite{CMS:2024nht} at $\sqrt{s}=13$ TeV, corresponding to an integrated luminosity of 138 fb$^{-1}$. Different coloured points represent $\Trh$-values required to satisfy the observed DM abundance for a DM of mass 1 MeV. In the left (right) panel  we fix $\tilde{k}=0.01~(0.1),\,m_r=1 {\rm ~TeV}$.}
    \label{fig:sigma_collider}
\end{figure*}

In Fig.~\ref{fig:sigma_collider}, we illustrate how the cross section into photonic final states varies with the graviton mass. For comparison with experimental data, we use the recent CMS result~\cite{CMS:2024nht} from the search for new physics in high-mass diphoton events in proton--proton collisions at a center-of-mass energy of 13~TeV. The dataset was collected during 2016--2018 with the CMS detector at the LHC and corresponds to an integrated luminosity of 138~fb$^{-1}$. The gray dashed curve represents the prediction of our model, obtained using the relation 
\begin{align}
    \tilde{k}\equiv\frac{k}{M_P}=\frac{m_{G_1}}{x_1\,\Lambda_r}\,.
\end{align}
By fixing $\tilde{k}$, one can vary both $m_{G_1}$ and $\Lambda_r$ over a suitable range and compare the resulting graviton-mediated $pp\to\gamma\gamma$ cross section with the CMS measurement (shown in black). The parton-level events are generated using CalcHEP~\cite{Pukhov:1999gg}, using the CTEQ6L distribution function. For each allowed pair $\left[m_{G_1},\,\Lambda_r\right]$, one can identify values of $\Trh$ that yield the correct relic abundance for a 1~MeV DM candidate; these values of $\Trh$ are indicated by the color scale. Interestingly, for $\tilde{k}=0.01$ (left panel), the present data exclude the region with $\Trh\lesssim 26~\text{GeV}$ and $m_{G_1}\lesssim 2230~\text{GeV}$, for a scalar DM of mass 1 MeV. Increasing $\tilde k=0.1$, as shown in the right panel, the constraints becomes more stringent, forbidding $m_{G_1}\lesssim4630$ GeV and corresponding $\Trh\lesssim$15.8 GeV. This figure highlights a striking complementarity between collider searches and early Universe cosmology. The key message is that collider measurements can impose stringent constraints not only on the mass scale and coupling of new physics, but also on the cosmological history of the early Universe.

\subsubsection{From reheating}
Away from the sudden decay approximation for reheating, the bath temperature may rise to a temperature $\Tmax\gg\Trh$~\cite{Giudice:2000ex,Kolb:2003ke,Rangarajan:2008zb}. It is plausible that the DM relic density may be established during this reheating period, in which case its abundance will significantly differ from freeze-in calculations assuming radiation domination. In particular, it has been observed that if the DM is produced during the transition from matter to radiation domination via an interaction rate that scales like $\gamma(T)\propto T^n$, for $n>12$ the DM abundance is enhanced by a boost factor proportional to $(\Tmax/\Trh)^{n-12}$~\cite{Garcia:2017tuj}, whereas for $n\leq 12$ the difference between the standard freeze-in calculation differ only by an $\mathcal{O}(1)$ factor from calculations taking into account non-instantaneous reheating. It has also been highlighted that the critical mass dimension of the operator at which the instantaneous decay approximation breaks down depend on the equation of state $\omega$, or equivalently, to the shape of the inflationary potential at the reheating epoch~\cite{Bernal:2019mhf,Garcia:2020eof,Co:2020xaf,Ahmed:2021fvt,Barman:2022tzk}. Therefore, the exponent of the boost factor becomes $(\Tmax/\Trh)^{n-n_c}$ with $n_c \equiv 6+2\, \left ( \frac{3-\omega}{1+\omega} \right ) $, showing a strong dependence on the equation of state~\cite{Bernal:2019mhf}. However, as the precise determination of such boost factors depends on the details of the reheating mechanism, in particular, the shape of the inflationary potential during reheating, it is beyond the scope of the present study. Given an inflationary model, a measurement (upper limit) on the tensor-to-scalar ratio can be translated into an upper bound on $\Trh$, thereby constraining the scale of inflation. Precision measurements of primordial element abundances from Big Bang nucleosynthesis (BBN) suggest that the reheating temperature, $\Trh$, must be at least a few MeV~\cite{Sarkar:1995dd, Kawasaki:2000en,Hannestad:2004px, DeBernardis:2008zz, deSalas:2015glj,Hasegawa:2019jsa}. On the other hand, typical inflationary models predict an upper bound on $\Trh$ around \( 10^{16}~\mathrm{GeV} \) (see, e.g., Ref.~\cite{Linde:1990flp}). However, a high reheating temperature can potentially lead to issues with long-lived exotic relics that risk overclosing the Universe. A well-known example of this is the cosmological gravitino problem in supergravity scenarios~\cite{Moroi:1993mb}. As a result, supergravity models usually impose an upper limit on the reheating temperature of around \( 10^{10}~\mathrm{GeV} \), with even stricter constraints if the gravitino is light. For the graviton and radion of masses $(\gtrsim\mathcal{O} (1\,\text{TeV}))$, both particles decay before the onset of BBN (approximately before one second), provided $\Lambda_r \lesssim 10^{10}\,\text{GeV}$.

A comprehensive analysis of inflation and the subsequent production of new physics after inflation lies beyond the scope of this work. Nevertheless, we note that a recent study~\cite{Bernal:2026oxa} investigated inflation within extra-dimensional frameworks from two complementary perspectives. The first is the so-called Dark Dimension (DD) scenario, which consists of a single flat extra dimension of size $L = 2\pi r_c$, where $r_c$ denotes the compactification radius, with a brane located at $y = 0$. The second framework is the Randall--Sundrum (RS) model, based on a five-dimensional anti-de Sitter geometry. Within this setup, both RS1 and RS2 realizations were examined: RS1 includes two branes at $y = 0$ and $y = \pi r_c$ (with the latter identified as the IR brane), while RS2 contains a brane at $y = 0$ and a second brane effectively pushed to infinity. The authors have also confirmed previous result, showing that 5D evolution leads to a correction of the Hubble parameter on the brane $H^2\propto\rho^2$~\cite{Binetruy:1999ut,Kanti:1999sz,Kraus:1999it} ($\rho$ being matter density on the brane), which is different from the 4D case where $H^2\propto\rho$. In these constructions, the inflaton field $\phi$, together with all SM fields, was assumed to be localized on a brane (specifically the IR brane in RS1). Two representative inflationary scenarios were then explored: monomial inflation and the simplest $\alpha$-attractor model. The authors showed that the monomial inflation scenario is disfavored by current ACT observations~\cite{AtacamaCosmologyTelescope:2025blo,AtacamaCosmologyTelescope:2025nti}, whereas the $\alpha$-attractor framework remains in good agreement with the data. In the RS1 setup, they also noted an important tension: matching cosmological observations requires an effective scale $\Lambda_r > 10^{13}\,\mathrm{GeV}$, which is vastly larger than the TeV scale typically needed to address the hierarchy problem. This suggests that, if the inflaton is confined to the IR brane, the RS1 framework may struggle to simultaneously explain both inflationary data and the hierarchy problem.

\subsubsection{From $\DNeff$}
For BBN to proceed successfully, the Universe must not contain a substantial abundance of additional relativistic degrees of freedom beyond those in the SM. Any such extra radiation energy density is conventionally parametrized through $\Delta N_{\rm eff}$, and its presence during the BBN epoch can modify the expansion rate, thereby impacting the formation of light elements. Consequently, both current and future constraints on $\DNeff$ from CMB, BBN, and their combination place stringent limits on scenarios that predict excess radiation prior to or during BBN. Here we consider a scenario where the decay of massive graviton (apart from the SM fields) produces massless dark radiation (DR) with spin-0, spin-1/2 and spin-1.

The effective number of neutrino species $N_\text{eff}$ is defined through the total radiation energy density in the late Universe (at a photon temperature $T_{\DNeff}$) via
\begin{eqnarray}
    \rho_\text{rad}(T_{\DNeff}) &=& \rho_\gamma + \rho_\nu + \rho_{\rm DR}
     \nonumber\\&=&\left[1 + \frac78 \left(\frac{T_\nu}{T_\gamma}\right)^4 N_\text{eff}\right]\rho_\gamma(T_{\DNeff})\,,
\end{eqnarray}
where $\rho_\gamma$, $\rho_\nu$, and $\rho_{\rm DR}$ denote the photon, SM neutrino, and dark radiation energy densities, respectively, and $T_\nu/T_\gamma = (4/11)^{1/3}$. The temperature $T_{\DNeff}$ specifies the epoch at which $N_\text{eff}$ is evaluated. Experimental bounds on $\DNeff$ are typically quoted at $T_{\DNeff} = T_\text{BBN}$ and $T_{\DNeff} = T_\text{CMB}$, with $T_\text{CMB}$ corresponding to photon decoupling. Within the SM, accounting for non-instantaneous neutrino decoupling yields $N_\text{eff}^\text{SM} = 3.044$~\cite{Dodelson:1992km, Hannestad:1995rs, Dolgov:1997mb, Mangano:2005cc, deSalas:2016ztq, EscuderoAbenza:2020cmq, Akita:2020szl, Froustey:2020mcq, Bennett:2020zkv}. In the presence of additional dark radiation, one finds
\begin{eqnarray}
 \label{eq:DNeff}
\DNeff &\equiv& N_\text{eff} - N_\text{eff}^\text{SM}
= \frac87 \left(\frac{11}{4}\right)^{4/3}\frac{\rDR(T_{\DNeff})}{\rho_\gamma(T_{\DNeff})}
\nonumber\\&=& \frac{43}{7}\left(\frac{11}{4}\right)^{4/3}
\left(\frac{\mathcal{B}}{1-\mathcal{B}}\right)
\left(\frac{43/4}{\gs(T_d)}\right)^{1/3}\,,
\end{eqnarray}
where $\mathcal{B}$ denotes the branching fraction of the graviton into DR of different spins. In deriving Eq.~\eqref{eq:DNeff} entropy conservation has been applied from decoupling temperature $T_d$ till $T_{\DNeff}$, and we have used $\gss\simeq\gs$ around $T_{\DNeff}$. The decoupling temperature of the decaying graviton satisfies
\begin{eqnarray}\label{eq:Td}
T_d&=&\left[\frac{3\,\Gamma_\mathscr{G}}{\pi}
\sqrt{\frac{10}{\gs(T_d)}}\,M_P\right]^{1/2}\nonumber\\&\simeq& 5.2\times 10^8\,\text{GeV}\,\left(\frac{m_G}{1\,\text{TeV}}\right)^{3/2}\,\left(\frac{10\,\text{TeV}}{\Lambda_r}\right)\,,
\end{eqnarray}
where $m_G$ is the mass of the decaying graviton. Using the expressions for graviton decay rates in Appendix.~\ref{sec:grav-decay} we find, 
\begin{align}
&\DNeff\simeq
\begin{dcases}
0.01\times\left(\frac{\mathcal{B}}{1/1095}\right) & \text{for spin-0}\,,
\\[10pt]
0.03\times\left(\frac{\mathcal{B}}{3/1097}\right) & \text{for spin-1/2}\,,
\\[10pt]
0.13\times\left(\frac{\mathcal{B}}{13/1107}\right) & \text{for spin-1}\,,
\\[10pt]
\end{dcases}
\end{align} 
ignoring masses of the SM particles and considering $\gs(T_d)=106$. This shows, spin-1 DR contributes maximally to $\DNeff$. 
\begin{figure*}[tbh]
    \centering        \includegraphics[scale=0.42]{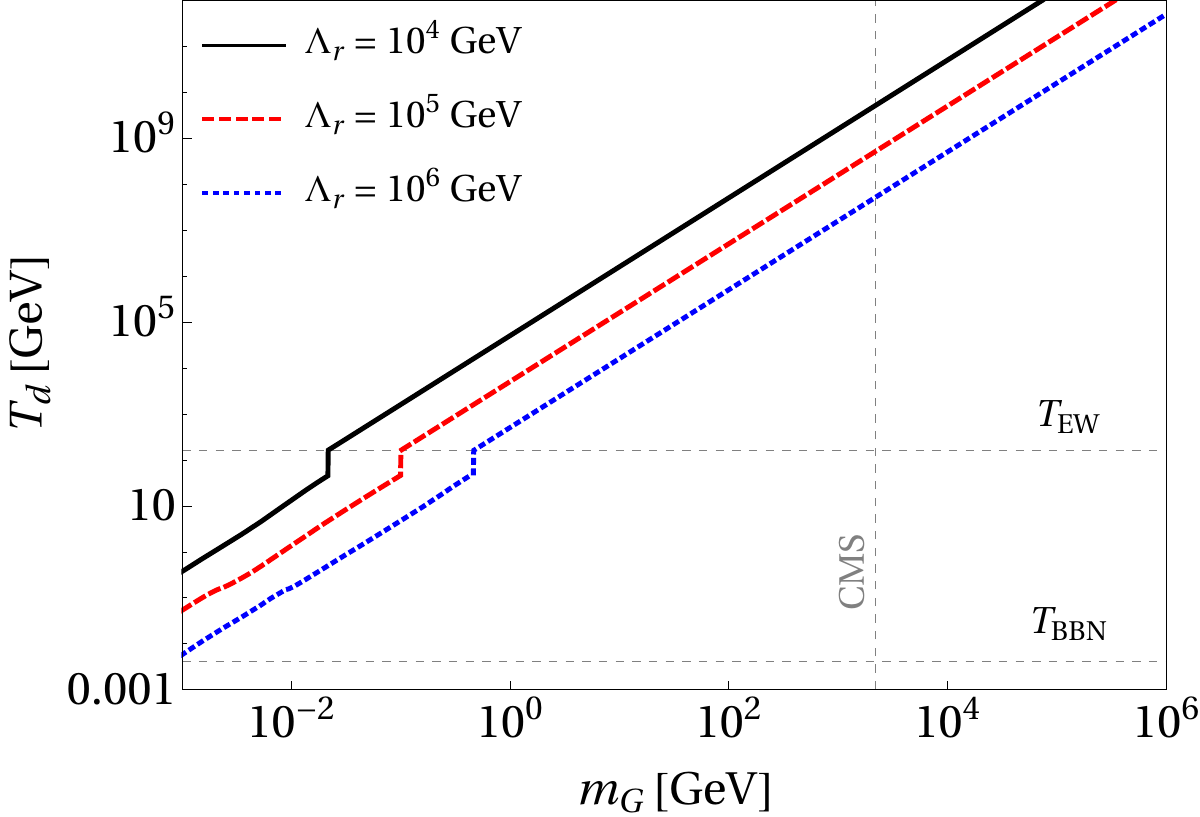}    \includegraphics[scale=0.42]{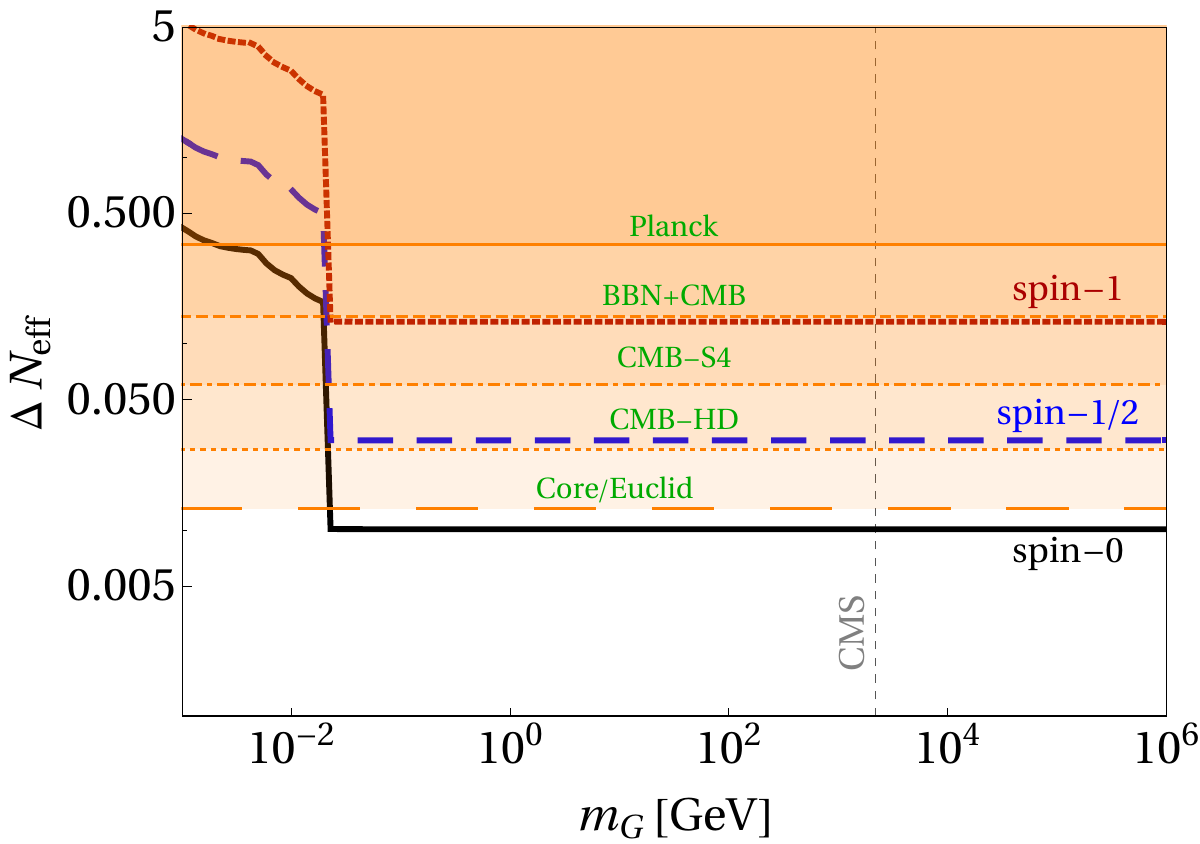}
    \caption{Left: Decoupling temperature $T_d$, as a function of graviton mass [cf.Eq.~\eqref{eq:Td}] for spin-1 DR. Different curves correspond to different choices of $\Lambda_r$. Right: Contribution to dark radiation from graviton decay, considering different spins, shown via different black curves, as a function of graviton mass, following Eq.~\eqref{eq:DNeff}. The shaded region is discarded from present (Planck) and future predictions of $\DNeff$ from different experiments. The vertical dashed line shows lower bound from CMS on first KK-graviton mass in RS model (see text for details).} 
    \label{fig:DR}
    \end{figure*}

Within the standard $\Lambda$CDM cosmological framework, the Planck legacy data yield
$N_{\rm eff} = 2.99 \pm 0.34$ at 95\%~CL~\cite{Planck:2018vyg}. Incorporating baryon acoustic
oscillation (BAO) measurements significantly sharpens this determination to
$N_{\rm eff} = 2.99 \pm 0.17$ at the $1\sigma$ level. Upcoming CMB surveys, such as SPT-3G~\cite{SPT-3G:2014dbx} and the Simons Observatory~\cite{SimonsObservatory:2018koc}, are
expected to improve upon Planck’s sensitivity. In particular, CMB-S4~\cite{Abazajian:2019eic}
and CMB-HD~\cite{CMB-HD:2022bsz} aim for remarkable precision, targeting sensitivities of
$\Delta N_{\rm eff} \simeq 0.06$ and $\Delta N_{\rm eff} \simeq 0.027$ (95\%~CL), respectively. A joint analysis combining BBN and CMB constraints yields
$N_{\rm eff} = 2.880 \pm 0.144$~\cite{Yeh:2022heq}. Looking ahead, next-generation
satellite missions such as COrE~\cite{COrE:2011bfs} and Euclid~\cite{EUCLID:2011zbd}
are projected to set limits at the level of $\Delta N_{\rm eff} \lesssim 0.013$ ($2\sigma$). In the right panel of Fig.~\ref{fig:DR}, we present the contribution to $\DNeff$ arising from the decay of a decoupled massive graviton into DR species with different spins. It is worth noting that current Planck constraints already exclude $m_G\lesssim 0.02$~GeV when spin-1 or spin-1/2 DR contributes to $\DNeff$, whereas the corresponding bound for spin-0 is much weaker, excluding only $m_G\lesssim 0.02$~MeV. Heavier gravitons remain unconstrained by existing Planck limits, but may be probed by the improved sensitivities of future experiments such as CMB-S4 and CMB-HD, particularly in the spin-1 and spin-1/2 cases. Spin-0 scenario, however, remains safe. Notably, present CMS analysis offers complementary coverage to CMB observations by ruling out $m_G\lesssim 2.2$~TeV, in the context of RS model. 

It is worth stressing that, although Eq.~\eqref{eq:DNeff} appears independent of $\Lambda_r$, as it depends solely on the branching fraction $\mathcal{B}$, a dependence on $\Lambda_r$ nonetheless enters through Eq.~\eqref{eq:Td}, as one can see from the left panel. For a fixed graviton mass, increasing $\Lambda_r$ suppresses the graviton decay width, thereby extending its lifetime. As a result, the graviton decays later, when the Hubble expansion rate has decreased to $H(T_d)\simeq \Gamma_\mathscr{G}$, corresponding to a cooler Universe. This is also evident from the left panel of Fig.~\ref{fig:DR}. The behavior of $\DNeff$ at small graviton masses can be directly understood from its dependence on the entropy degrees of freedom at the decoupling temperature, as seen in Eq.~\eqref{eq:DNeff}. For fixed $\Lambda_r$, the decoupling temperature scales as $T_d \propto m_G^{3/2}$, implying that lighter gravitons decouple at later times when the temperature of the thermal bath is lower. Since the effective entropy degrees of freedom $\gs(T)$ decrease significantly as the Universe cools (notably across the QCD and electron--positron annihilation epochs), a smaller $T_d$ corresponds to a smaller value of $\gs(T_d)$. Consequently, the dilution of the DR component due to subsequent entropy injection into the SM plasma is reduced. This is reflected in Eq.~\eqref{eq:DNeff} through the factor $\gs(T_d)^{-1/3}$, leading to an enhanced contribution to $\DNeff$ for lighter gravitons. In contrast, heavier gravitons decouple earlier, when $\gs(T_d)$ is large, resulting in stronger entropy dilution and hence a suppressed $\DNeff$. This explains the rise of $\DNeff$ toward smaller graviton masses observed in. In summary, a massive graviton with $m_G\gtrsim 2$~TeV remains safe from current $\DNeff$ limits from Planck, considering its decays into dark radiation having intrinsic different spins.
\section{Implications for baryogenesis via leptogenesis}
\label{sec:lepto}
As advocated in the beginning, the goal of the present study is to address not only the DM abundance, but also the asymmetry in the brayonic sector. In order to achieve the latter we adopt the leptogenesis route to baryogenesis. Following the standard leptogenesis scenario~\cite{Yanagida:1979as,Davidson:2000er}, we introduce three SM gauge singlet RHNs $N_i$ (with $i= 1$, 2, 3) with the SM fields, that leads to an interaction Lagrangian,
\begin{equation} \label{eq:RHN-lgrng}
\mathcal{L} \supset -\frac12\, M_N\, \overline{N^c}\, N - y_N\, \overline{N}\, \widetilde{H}^\dagger\, \psi_L + {\rm H.c.}\,,
\end{equation}   
where RHNs are assumed to be mass diagonal, and all the generational indices are suppressed. The first term is the lepton number violating Majorana mass term for RHN. The second piece corresponding to the interaction Lagrangian is important since it is responsible for the generation of active neutrino mass, as well as the lepton asymmetry via CP-violating decay of the RHNs, which eventually gets converted into the observed baryon asymmetry via sphalerons. Here $\widetilde{H}=i\,\sigma_2\,H^\star$, where $\sigma_2$ is the Pauli spin matrix. With the Lagrangian in Eq.~\eqref{eq:RHN-lgrng}, we can write down the subsequent action as,
\begin{eqnarray}
\mathcal{S}_{\text{rad}-N}&\supset&\,-\int\,d^4x\,\left(\frac{4r}{\sqrt{6}\,\Lambda_r}\right)\,\left[\frac{1}{2}\,M_N\,\overline{N^c}\,N+\right.\nonumber\\&&\left.y_{N}\,\overline{N}\widetilde{H}^{\dagger}\psi_L+\text{H.c.}\right]\,.
\end{eqnarray}
Once again, we rescale the RHN mass $M_N\to M_N^0\,e^{-k\,r_c\,\pi}$. Consequently, for $kr_c\sim\mathcal{O}(10)$, we have TeV-scale RHN for $M_N^0\sim\mathcal{O}(M_P)$.
\subsection{Neutrino mass and CP-asymmetry}
Once the neutral component of $H$ acquires a nonzero VEV, the neutrino Dirac mass term can be written as  
\begin{align}
m_D=\frac{y_N}{\sqrt{2}}v\,,
\end{align}  
following Eq.~\eqref{eq:RHN-lgrng}. Together with the bare Majorana mass of the RHNs, this generates light neutrino masses through the Type-I seesaw mechanism~\cite{Gell-Mann:1979vob, Mohapatra:1979ia},  
\begin{align}
m_{\nu}\simeq -m_D\,M_N^{-1}\,m_D^T\,,
\label{NM}
\end{align}  
assuming $m_D \ll M_N$. Diagonalization yields  
\begin{align}
m_{\nu}= \mathcal{U}^*\,m_{\nu}^d\,\mathcal{U}^{\dagger}\,,
\end{align}  
where $m_{\nu}^d={\rm diag}(m_1,\,m_2,\,m_3)$ and $\mathcal{U}$ is the PMNS matrix~\cite{ParticleDataGroup:2022pth}, with the charged lepton mass matrix taken diagonal. To generate the complex Yukawa structure required for CP-violating RHN decays, we employ the Casas–Ibarra (CI) parametrization~\cite{Casas:2001sr},  
\begin{align}
y_N = \frac{\sqrt{2}}{v}\,\mathcal{U}\,\sqrt{m_{\nu}^d}\,\mathbb{R}^T\,\sqrt{M_N^d}\,,
\label{CI}
\end{align}  
where $M_N^d$ is the diagonal RHN mass matrix, $\mathbb{R}$ is a complex orthogonal matrix, $\mathbb{R}^T\mathbb{R}=I$, chosen as  
\begin{align}
\mathbb{R} =
\begin{pmatrix}
0 & \cos\alpha & \sin\alpha\\
0 & -\sin\alpha & \cos\alpha\\
1 & 0 & 0
\end{pmatrix}\,,
\label{eq:rot-mat}
\end{align}  
with $\alpha=a+ib$ a complex angle, and we have considered the RHNs to be mass diagonal. The mass matrix $m_{\nu}^d$ is determined from the latest oscillation data~\cite{ParticleDataGroup:2020ssz,ParticleDataGroup:2022pth}. Note that, in writing the above rotation matrix, we have considered $N_3$ to be extremely heavy and decoupled, consequently only $N_{1,2}$ take part in leptogenesis. Consequently, the lightest active neutrino mass can be set to zero. 
The CP asymmetry parameter is defined as  
\begin{align}
\epsilon_i = \frac{\sum_j\Gamma(N_i\to\ell_j H)-\sum_j\Gamma(N_i\to\overline{\ell_j}\,\overline{H})}{\sum_j\Gamma(N_i\to\ell_j H)+\sum_j\Gamma(N_i\to\overline{\ell_j}\,\overline{H})}\,.
\end{align}  

As the present set-up naturally forbids high-scale leptogenesis (to consistently address the hierarchy problem) and gives rise to TeV-scale RHNs, hence in order to satisfy the observed baryon asymmetry we would require to resonantly enhance the CP-asymmetry leading to resonant leptogenesis. A resonant enhancement of the CP asymmetry in $N_1$ decay occurs when the mass difference between $N_1$
and $N_2$ is of the order of the decay widths. The CP-asymmetry due to $N_1$ decay then reads~\cite{Pilaftsis:2005rv,Pilaftsis:2003gt,Anisimov:2005hr,Davidson:2008bu},
\begin{equation}
\epsilon_1^{\rm reso} =
\frac{\mathrm{Im}\!\left[\,(Y_N^\dagger Y_N)_{12}^2\,\right]}
{(Y_N^\dagger Y_\nu)_{11}\,(Y_N^\dagger Y_\nu)_{22}}
\cdot
\frac{(M_1^2 - M_2^2)\,M_1\,\Gamma_2}
{(M_1^2 - M_2^2)^2 + M_2^2 \Gamma_2^2}\,,
\end{equation}
where we have neglected the flavor effects. This result is only applicable to the case of two nearly
degenerate heavy Majorana neutrinos. It is worth remarking that two Majorana neutrinos are suﬃcient for generating the light neutrino masses and explaining the baryon number asymmetry.
\subsection{Generation of baryon asymmetry}
To track the evolution of $N_1$, along with the B$-$L charge with temperature of the SM bath, we solve the following set of coupled Boltzmann equations,
\begin{eqnarray}\label{eq:beq-lepto}
\frac{dY_r}{dz}&=&-\frac{1}{z\,\mathcal{H}}\,\langle\Gamma_{r\to\text{SM}\,\text{SM}}\rangle\,\left(Y_r-Y_r^{\rm eq}\right)-\nonumber\\&&\frac{s}{z\,\mathcal{H}}\,\langle\sigma v\rangle_{\text{SM}\, \text{SM}\to rr}\,\left(Y_r^2-Y_{r,\text{eq}}^2\right)\,,
\nonumber\\
\frac{dY_{\mathcal{G}}}{dz}&=&-\frac{1}{z\,\mathcal{H}}\,\langle\Gamma_{\mathcal{G}\to \text{SM}\,\text{SM}}\rangle\,\left(Y_{\mathcal{G}}-Y_\mathcal{G}^{\rm eq}\right)-\nonumber\\&&\frac{s}{z\,\mathcal{H}} \,\langle\sigma v\rangle_{\text{SM}\, \text{SM}\to\mathcal{G}\mathcal{G}}\,\left(Y_\mathcal{G}^2-Y_{\mathcal{G},\text{eq}}^2\right)\,,
\nonumber\\
\frac{dY_{N_1}}{dz}&=&\frac{1}{z\,\mathcal{H}}\,\langle\Gamma_{\mathcal{G}\to N_1\,N_1}\rangle\,Y_{\mathcal{G}}+\frac{1}{z\,\mathcal{H}}\,\langle\Gamma_{r\to N_1\,N_1}\rangle\,Y_r-\nonumber\\&&\frac{s}{z\,\mathcal{H}}\,\langle\sigma v\rangle_{\text{SM}\, \text{SM}\to N_1N_1}\,\left(Y_{N_1}^2-Y_{N_1,\text{eq}}^2\right)\,,
\nonumber\\&&
-\frac{1}{z\,\mathcal{H}}\,\langle\Gamma_{N_1}\rangle\left(Y_{N_1}-Y_{N_1}^{\rm eq}\right)\,,
\nonumber \\
\frac{dY_{B-L}}{dz}&=&\frac{\langle\Gamma_{N_1}\rangle}{z\,\mathcal{H}}\,\epsilon^{\rm reso}_1\,\left(Y_{N_1}-Y_{N_1}^{\rm eq}\right)
-\nonumber\\&&\frac{\langle\Gamma_{N_1}\rangle}{\mathcal{H}}
\frac{z}{4}\,\frac{\widetilde m_1}{m_\star}\,K_1(z)\,Y_{B-L}\,,
\end{eqnarray}
where $z=M_1/T$. Here, $
\widetilde m_1=\left(m_D^\dagger\,m_D\right)_{22}/M_1\approx m_{\nu,1}^2/M_1$ and and $m_\star\simeq 10^{-3}$ eV is the equilibrium neutrino mass~\cite{Buchmuller:2004nz}. The thermally averaged decay width reads,
\begin{align}
& \langle \Gamma_i\rangle=\frac{K_1(M_i/T)}{K_2(M_i/T)}\times\Gamma_{i\to jj}\,, 
\end{align}
where $i$ and $j$ are the decaying particle and the decay products, respectively. The first and second line of Eq.~\eqref{eq:beq-lepto} corresponds to the yield of frozen-in radions and KK-gravitons, respectively, that are created (on-shell) either from the inverse decays (2-to-1 process) or from the scattering (2-to-2 process) of the bath particles, where the latter is suppressed by $1/\Lambda_r^4$, and hence sub-dominant in presence of decay. The third line takes care of the evolution of RHN yield, where we include RHN production from graviton and radion decays, as well as graviton and radion mediated scattering (included in $\langle\sigma v\rangle_{\text{SM}\,\text{SM}\to N_1\,N_1}$), on top of usual production via inverse decay of the bath particles $(\ell\,H\to N_1)$. The last line of equation determines the net asymmetry produced. Sphaleron interactions are in equilibrium in the temperature range between $\sim$ 100 GeV and $10^{12}$ GeV, and they convert a fraction of a non-zero B$-$L asymmetry into a baryon asymmetry via
\begin{align}
Y_B\simeq a_{\rm sph}\,Y_{B-L}=\frac{8\,N_F+4\,N_H}{22\,N_F+13\,N_H}\,Y_{B-L}\,,    
\end{align}
where $N_F$ is the number of fermion generations and $N_H$ is the number of Higgs doublets, which in our case: $N_F = 3,\,N_H = 1$ and $a_{\rm sph}\simeq 28/79$. In leptogenesis, where purely a lepton asymmetry is generated, $B-L=-L$. This is converted into the baryon asymmetry via sphaleron transition~\cite{Buchmuller:2004nz}. Finally, the observed baryon asymmetry of the Universe is given by $Y_B^0 \simeq 8.75\times 10^{-11}$. 

\begin{figure*}[tbh]
    \centering    
    \includegraphics[scale=0.42]{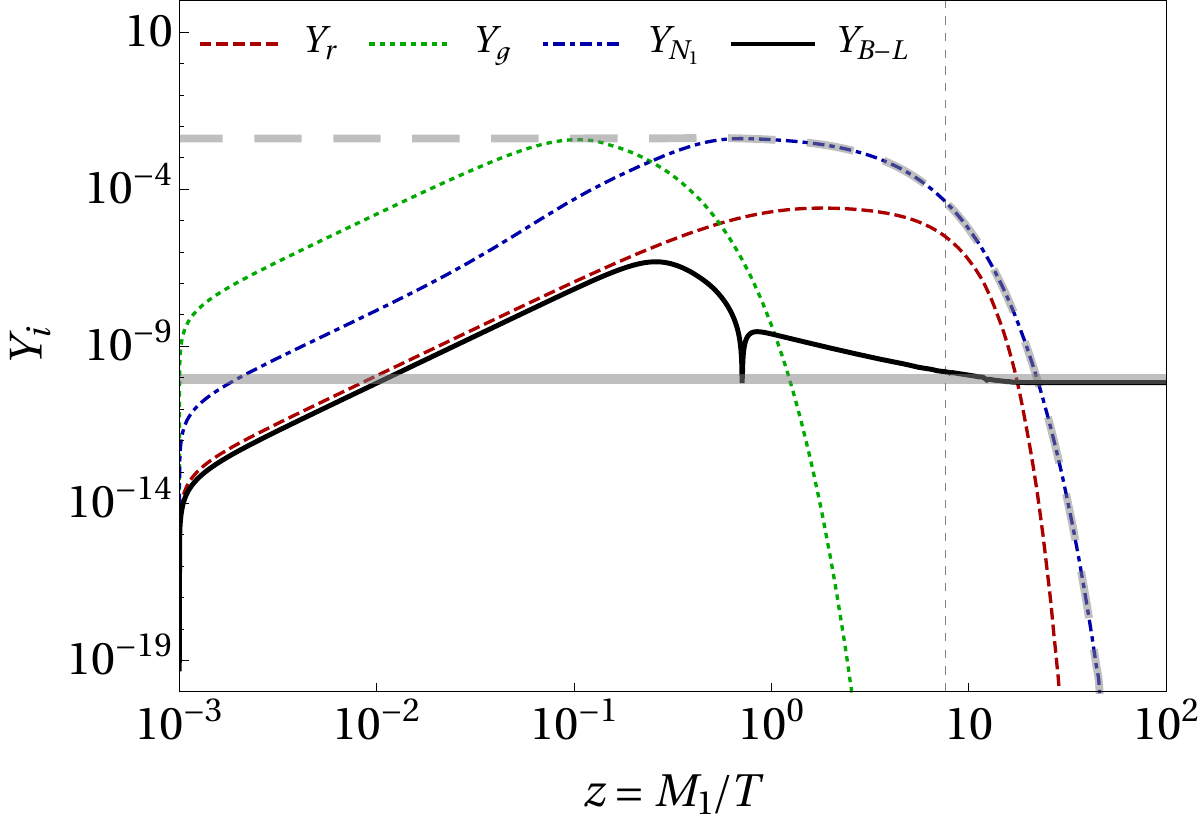}~\includegraphics[scale=0.42]{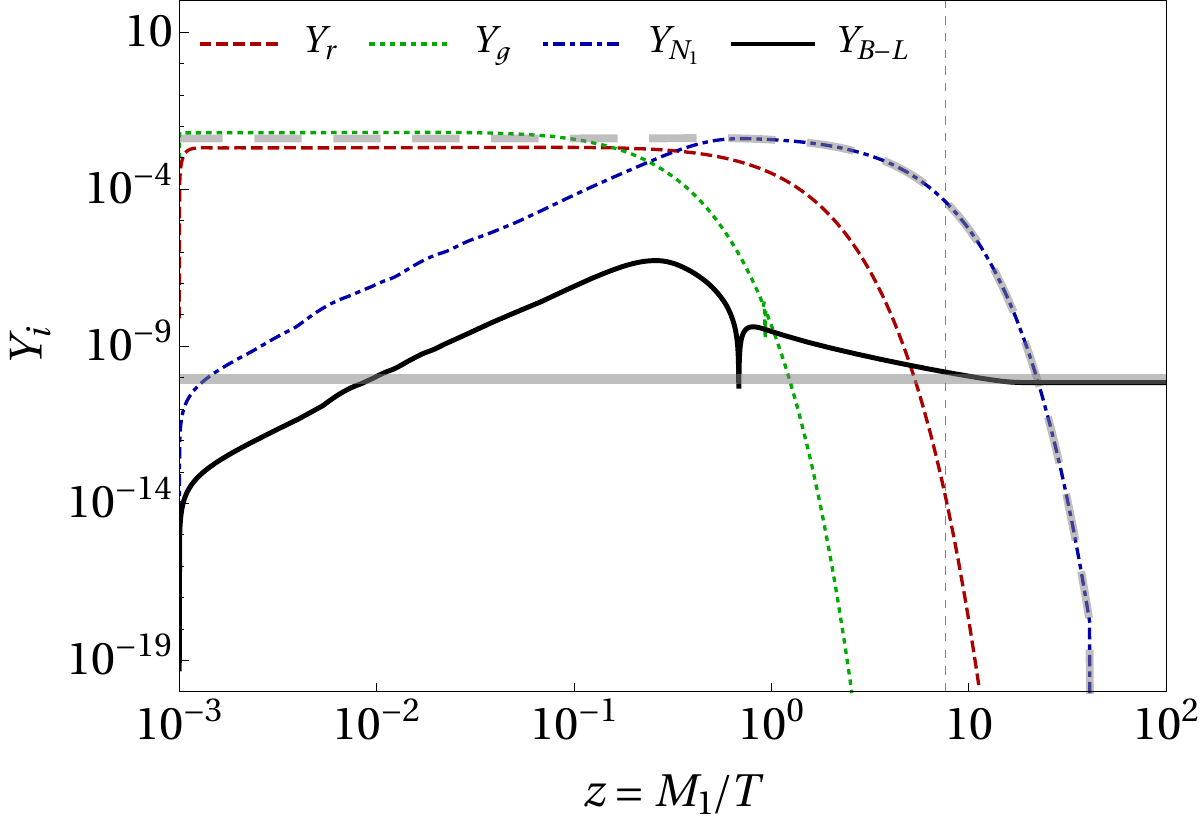}
    \caption{Evolution of radion, graviton, RHN and B$-$L yield, shown via red-dashed, green-dotted, blue-dotdashed and black solid curves, respectively. In the left panel we have chosen $\Lambda_r=3\times10^{10}$ GeV, while $\Lambda_r=10^5$ GeV in the right panel. In the left panel we also show the thermal RHN yield via black dashed curve (on top of the blue dot-dahsed cuve). In all cases we have fixed $m_r=2$ TeV and $M_1=500$ GeV, and the graviton mass follows from Tab.~\ref{tab:mn}. The vertical dashed gray line corresponds to the temperature of sphaleron transition. The horizontal gray line represents the value of the observed baryon asymmetry $Y_B^0 \simeq 8.75\times 10^{-11}$.}
    \label{fig:lepto-yld}
\end{figure*}

We present the numerical solution to Eq.~\eqref{eq:beq-lepto} in Fig.~\ref{fig:lepto-yld}, which shows the evolution of the radion, graviton, RHN, and the absolute value of the asymmetry as functions of $z$. The RHN mass is fixed to be 500 GeV, considering $kr_c=10$ such that the hierarchy problem is once again addressed. The radion mass is fixed to 2 TeV, satisfying the collider limit and we consider the contribution from the lowest KK-graviton mass mode, following Tab.~\ref{tab:mn}. As the effective scale $\Lambda_r$ decreases, the yields of the radion and graviton increasingly approach their equilibrium distributions, as evident from the comparison between the left and right panels. This behavior is expected, since a smaller $\Lambda_r$ enhances the interaction rate, allowing it to compete effectively with the Hubble expansion rate. The RHN yield gradually builds up from a negligible initial value, and its subsequent decay generates the final B$-$L asymmetry, represented by the solid black curve. The CI parameters are chosen such that the resulting asymmetry reproduces the observed value while maintaining perturbativity of the Yukawa coupling. For TeV-scale leptogenesis, it is essential to account for flavour effects~\cite{Davidson:2008bu}, where the dynamics of individual lepton flavours play a significant role in generating the final asymmetry. Since the framework of flavoured leptogenesis is extensively developed in the literature and applies to our scenario without modification, we do not elaborate on it here. Instead, we highlight that the present set-up naturally accommodates low-scale leptogenesis, placing it within the potential reach of collider experiments (see, for example, Refs.~\cite{Atre:2009rg,Drewes:2016jae,Antusch:2017pkq,Chakraborty:2022pcc}). 

{\it Baryon-dark matter coincidence:} We conclude by reflecting on the viability of explaining the baryon--DM coincidence within this framework. Since the baryon asymmetry is predominantly generated by RHNs originating from the thermal bath, their efficient production in the early Universe requires the reheating temperature to satisfy $\Trh \gtrsim M_1$. Consequently, a successful realization of leptogenesis driven by the CP-violating decays of a $500\ \text{GeV}$ RHN necessitates $\Trh \gtrsim 500\,\text{GeV}$. As demonstrated in the right panel of Fig.~\ref{fig:2}, the observed dark matter relic abundance can be consistently reproduced for $\Trh \gtrsim 500\,\text{GeV}$ by appropriately adjusting the effective scale $\Lambda_r$. Hence, for a moderately high reheating temperature, our setup naturally accommodates the baryon--DM coincidence while simultaneously offering a coherent resolution to the hierarchy problem. 
\section{Conclusions}
\label{sec:concl}
Within the framework of the RS setup, we have explored the production of new physics states. In our scenario, all SM fields are confined to the IR brane, while only the spin-2 graviton propagates in the bulk. The RS construction additionally predicts a scalar degree of freedom, the radion. Building on this setup, we extend the SM by introducing gauge-singlet fields that can either serve as dark matter (DM) candidates or generate the baryon asymmetry through standard leptogenesis. Even without direct couplings between DM and the SM, the DM production is inevitably facilitated by their gravitational interactions mediated by the graviton and the radion. Interestingly, satisfying all collider constraints from LHC and cosmological bounds, e.g., from reheating and from $\DNeff$, the observed DM relic abundance can be fully explained while simultaneously addressing the hierarchy problem. We show that the reheat temperature is sufficient enough to produce the RHNs abundantly. These RHNs subsequently undergo CP-violating, out-of-equilibrium decays, yielding the observed matter–antimatter asymmetry of the Universe. Since the RHN mass is constrained to be at the TeV scale in order to solve the hierarchy problem, achieving a sufficient CP asymmetry necessitates resonant enhancement. We further examine the impact of collider constraints within this framework, with particular emphasis on their implications for early Universe cosmology. In particular, existing bounds from LHC searches for gravitons in di-photon final states impose significant restrictions on the reheating temperature, depending on the graviton mass and size of the effective scale. In summary, the RS framework, together with a simple extension of the SM particle content, offers a unified and coherent resolution to the coincidence problem while consistently addressing the hierarchy issue.
\section*{Acknowledgment}
BB would like to acknowledge  the hospitality at IIT, Hyderabad, during the Phoenix-2025 conference, during which this work was initiated. P.K.P. acknowledges the Ministry of Education, Government of India, for providing financial support for his research via the Prime Minister’s Research Fellowship (PMRF) scheme. We would like to thank the anonymous referee for many useful suggestions.
\onecolumngrid
\appendix
\section{Radion decay rates}
\label{sec:rad-decay}
To obtain relevant radion-matter vertices, we implemented the model in LanHEP~\cite{Semenov:2008jy} and extracted the decay widths as well as cross-sections utilizing CalcHEP~\cite{Pukhov:1999gg}. 
\begin{table}[htb!]
    \centering
    \renewcommand{\arraystretch}{1.5} 
    \begin{tabular}{c|c}
        \hline
        Interactions & Vertices \\
        \hline\hline
        $r\,H(p_1)\,H(p_2)$ & $(i/\Lambda_r)\,(2\,p_1\cdot p_2+4\,m_h^2)$ \\
        \hline
        $r\,V(p_1)\,V(p_2)$ & $(-2\,i\,m_V^2/\Lambda_r)\,\eta_{\alpha\beta}$
        \\
        \hline        $r\,f(p_1)\,\bar{f}(p_2)$ & $i\,m_f/\Lambda_r$\\
      \hline
    \end{tabular}
    \caption{3-legged vertices for the radion-matter interactions in unitary gauge. Here, $m_{h,V,f}$ corresponds to the Higgs, massive gauge boson and the SM fermion masses, respectively.}
    \label{tab:vertexRad}
\end{table}
\subsection{To Standard Model}
\subsubsection{Before EWSB}
Before EWSB, the only decay mode that is accessible to the radion is the SM Higgs, with a decay rate
\begin{align}
\Gamma_{r\to hh}=\frac{m_r^3}{768\,\pi\Lambda_r^2}\,.
\end{align}
\subsubsection{After EWSB}
After the EWSB, the radion can decay to SM fermions, Gauge bosons, and Higgs in the SM. The decay widths of these modes are given as follows,
\begin{align}
&\Gamma_{r\to\text{SM}\,\text{SM}}=\frac{m_r^3}{384\,\pi\Lambda_r^2}
\begin{dcases}
\sqrt{1-x_W}\left(1-x_W+\frac{3}{4}x_W^2\right)\,, 
\\[10pt]
\frac{1}{2}\,\sqrt{1-x_Z}\left(1-x_Z+\frac{3}{4}x_Z^2\right)\,, 
\\[10pt]
\frac{1}{2}\,\sqrt{1-x_h}\,,
\\[10pt]
2\,N_c\,x_f\,\left(1-x_f\right)^{3/2}\,,
\end{dcases}
\end{align}
where $x_i\equiv 4 m_i^2 / m_r^2$, $N_c=3(1)$ represents the colour factor for quarks (leptons) and $m_f$ is the mass of the SM fermions. 
\subsection{To dark matter}
\begin{align}
& \Gamma_{r\to\text{DM}\,\text{DM}}=\frac{m_r^3}{768\,\pi\Lambda_r^2}
\begin{dcases}
(1-x_{\rm DM})^{5/2} & \text{spin-0}\,,
\\[10pt]
8\,x_{\rm DM}\,(1-x_{\rm DM})^{3/2} & \text{spin-1/2}\,,
\\[10pt]
\sqrt{1-x_{\rm DM}}\,(1-4x_{\rm DM}+12x_{\rm DM}^2) & \text{spin-1}\,.
\end{dcases}
\end{align}
where $x_{\rm DM}\equiv4m_{\rm DM}^2/m_r^2$.
\section{Radion-mediated scattering cross-sections}
\label{sec:rad-med-cs}
\subsection{Before EWSB}
In this case, only $hh\to$ DM\,DM channel survives with the following production cross-sections,
\begin{align}
&\sigma_{hh\to\text{DM}\,\text{DM}}=
\frac{s}{18432\,\pi \Lambda_r^4}\,\sqrt{1-\frac{4\mdm^2}{s}}
\begin{dcases}
\frac{(s-4m_{\rm DM}^2)^2}{(s-m_r^2)^2+\Gamma_r^2\,m_r^2}\,, & \text{spin-0 DM}\,,
\\[10pt]
\frac{s^2-4s\,m_{\rm DM}^2+12\,\mdm^4}{(s-m_r^2)^2+\Gamma_r^2\,m_r^2}\,, & \text{spin-1 DM}\,,
\\[10pt]
\frac{32\,m_{\rm DM}^2\,\left(s-4m_{\rm DM}^2\right)}{(s-m_r^2)^2+\Gamma_r^2\,m_r^2}\,, & \text{spin-1/2 DM}\,.
\end{dcases}
\end{align}
\subsection{After EWSB}
Once the EW symmetry is broken, all the SM states becomes massive. Below we report all the production cross-sections for different DM spins. In the following we consider $x=2m_{\rm DM}/\sqrt{s},y=2m_{\rm SM}/\sqrt{s}$
\subsubsection{Spin-0 DM}
\begin{align}
& \sigma_{ll\to SS}=\frac{1}{36864\,\pi \Lambda_r^4}\frac{s^3y^2}{\left[(s-m_r^2)^2+\Gamma_r^2\,m_r^2\right]}(1-x^2)^{5/2}(1-y^2)^{1/2}\,,
\\&
\sigma_{qq\to SS}=\frac{1}{110592\,\pi \Lambda_r^4}\frac{s^3y^2}{\left[(s-m_r^2)^2+\Gamma_r^2\,m_r^2\right]}(1-x^2)^{5/2}(1-y^2)^{1/2}\,,
\\&
\sigma_{ZZ\to SS}=\frac{1}{663552\,\pi \Lambda_r^4\,}\,\frac{s^3(4-4y^2+3y^4)}{\left[(s-m_r^2)^2+\Gamma_r^2\,m_r^2\right]}(1-x^2)^{5/2}(1-y^2)^{-1/2}\,,
\\&
\sigma_{WW\to SS}=\frac{1}{663552\,\pi \Lambda_r^4\,}\,\frac{s^3(4-4y^2+3y^4)}{\left[(s-m_r^2)^2+\Gamma_r^2\,m_r^2\right]}(1-x^2)^{5/2}(1-y^2)^{-1/2}\,,
\\&
\sigma_{hh\to SS}=\frac{1}{18432\,\pi \Lambda_r^4\,}\,\frac{s^3}{\left[(s-m_r^2)^2+\Gamma_r^2\,m_r^2\right]}(1-x^2)^{5/2}(1-y^2)^{-1/2}\,.
\end{align}
\subsubsection{Spin-1 DM}
\begin{align}
& 
\sigma_{\ell\ell\to XX}=\frac{1}{147456\,\pi \Lambda_r^4}\frac{s^3(4-4x^2+3x^4)y^2}{\left[(s-m_r^2)^2+\Gamma_r^2\,m_r^2\right]}(1-x^2)^{1/2}(1-y^2)^{1/2}\,,
\\&
\sigma_{qq\to XX}=\frac{1}{442368\,\pi \Lambda_r^4}\frac{s^3(4-4x^2+3x^4)y^2}{\left[(s-m_r^2)^2+\Gamma_r^2\,m_r^2\right]}(1-x^2)^{1/2}(1-y^2)^{1/2}\,,
\\&
\sigma_{ZZ\to XX}=\frac{1}{2654208\,\pi\,\Lambda_r^4}\frac{s^3(4-4x^2+3x^4)(4-4y^2+3y^4)}{\left[(s-m_r^2)^2+\Gamma_r^2\,m_r^2\right]}(1-x^2)^{1/2}(1-y^2)^{-1/2} \,,
\\&
\sigma_{WW\to XX}=\frac{1}{2654208\,\pi\,\Lambda_r^4}\frac{s^3(4-4x^2+3x^4)(4-4y^2+3y^4)}{\left[(s-m_r^2)^2+\Gamma_r^2\,m_r^2\right]}(1-x^2)^{1/2}(1-y^2)^{-1/2}\,,
\\&
\sigma_{hh\to XX}=\frac{1}{73728\,\pi\,\Lambda_r^4}\frac{s^3(4-8x^2+7x^4-3x^6)}{\left[(s-m_r^2)^2+\Gamma_r^2\,m_r^2\right]}(1-x^2)^{-1/2}(1-y^2)^{-1/2}\,.
\end{align}
\subsubsection{Spin-1/2 DM}
\begin{align}
& \sigma_{\ell\ell\to NN}=\frac{1}{4608\,\pi \Lambda_r^4}\frac{s^3x^2y^2}{\left[(s-m_r^2)^2+\Gamma_r^2\,m_r^2\right]}(1-x^2)^{3/2}(1-y^2)^{1/2}\,,
\\&
\sigma_{qq\to NN}=\frac{1}{13824\,\pi \Lambda_r^4}\frac{s^3x^2y^2}{\left[(s-m_r^2)^2+\Gamma_r^2\,m_r^2\right]}(1-x^2)^{3/2}(1-y^2)^{1/2}\,,
\\&
\sigma_{ZZ\to NN}=\frac{1}{2304\,\pi \Lambda_r^4\,}\,\frac{s^3x^2(4-4y^2+3y^4)}{\left[(s-m_r^2)^2+\Gamma_r^2\,m_r^2\right]}(1-x^2)^{3/2}(1-y^2)^{-1/2}\,,
\\&
\sigma_{WW\to NN}=\frac{1}{82944\,\pi \Lambda_r^4\,}\,\frac{s^3x^2(4-4y^2+3y^4)}{\left[(s-m_r^2)^2+\Gamma_r^2\,m_r^2\right]}(1-x^2)^{3/2}(1-y^2)^{-1/2}\,,
\\&
\sigma_{hh\to NN}=\frac{1}{2304\,\pi \Lambda_r^4\,}\,\frac{s^3x^2}{\left[(s-m_r^2)^2+\Gamma_r^2\,m_r^2\right]}(1-x^2)^{3/2}(1-y^2)^{-1/2}\,.
\end{align}
\section{Graviton decay rates}
\label{sec:grav-decay}
The propagator of the $n^{\rm th}$ KK graviton mode, characterized by mass $m_{G_n}$, decay width $\Gamma_n$, and four-momentum $k$, can be written in the unitary gauge as  
\begin{align}
i \Delta^{\mathcal{G}}_{\mu\nu\alpha\beta}(k) = 
\frac{i P_{\mu\nu\alpha\beta}(k,m_{G_n})}{k^2 - m_{G_n}^2 + i m_{G_n} \Gamma_n} \,,
\label{eq:grav-prop}
\end{align}
where $P_{\mu\nu\alpha\beta}$ denotes the polarization sum of the graviton, expressed in terms of the polarization tensors $\epsilon^{s}_{\mu\nu}(k)$ for spin $s$:  
\begin{align}
P_{\mu\nu\alpha\beta}(k,m_{G_n})=\sum_s \epsilon^{s}_{\mu\nu}(k)\, \epsilon^{s}_{\alpha\beta}(k)=\frac{1}{2}\left(\mathcal{G}_{\mu\alpha} \mathcal{G}_{\nu\beta} + \mathcal{G}_{\nu\alpha} \mathcal{G}_{\mu\beta} 
- \frac{2}{3} \mathcal{G}_{\mu\nu} \mathcal{G}_{\alpha\beta}\right)\,.
\label{eq:grav-pol}
\end{align}
Here, the tensor $\mathcal{G}_{\mu\nu}$ is defined as  
\begin{align}
\mathcal{G}_{\mu\nu} \equiv \eta_{\mu\nu} - \frac{k_\mu k_\nu}{m_{G_n}^2} \,.
\end{align}
For an on-shell graviton, the polarization tensor $P_{\mu\nu\alpha\beta}$ satisfies 
\begin{align}
\eta^{\alpha\beta} P_{\mu\nu\alpha\beta}(k,m_{G_n}) 
= \eta^{\nu\mu} P_{\mu\nu\alpha\beta}(k,m_{G_n}) = 0 \,, 
\label{eq:A4}
\end{align}
and  
\begin{align}
k^\alpha P_{\mu\nu\alpha\beta}(k,m_{G_n}) 
= k^\beta P_{\mu\nu\alpha\beta}(k,m_{G_n}) 
= k^\mu P_{\mu\nu\alpha\beta}(k,m_{G_n}) 
= k^\nu P_{\mu\nu\alpha\beta}(k,m_{G_n}) = 0 \,.
\end{align}

For extracting relevant graviton-matter vertices, we have utilized FeynRules~\cite{Alloul:2013bka}. In Tab.~\ref{tab:vertex} we have tabulated the graviton-SM-SM vertices extracted from FeynRules. In all cases all the momenta $p_{i,j}$ are incoming (towards the vertex), following the FeynRules convention. We have defined,
\begin{align*}
& \mathcal{C}_{\mu\nu,\alpha\beta}=\Big[
\eta_{\alpha\beta}(p_{i\mu}\,p_{j\nu} + p_{i\nu}\,p_{j\mu})
- \eta_{\mu\nu}(p_{i\alpha}\,p_{j\beta})
+ \eta_{\alpha\mu}p_{i\nu}\,p_{j\beta}
+ \eta_{\alpha\nu}\,p_{i\mu}p_{j\beta}
\nonumber\\&
+ \eta_{\beta\mu}\,p_{i\alpha}\,p_{j\nu}
+ \eta_{\beta\nu}\,p_{i\alpha}p_{j\mu}
\Big]\,,    
\end{align*}
along with,
\begin{align*}
\mathcal{D}_{\mu\nu,\alpha\beta}=\eta_{\alpha\mu}\eta_{\beta\nu}+\eta_{\alpha\nu}\eta_{\beta\mu}- \eta_{\alpha\beta}\eta_{\mu\nu}\,.  
\end{align*}
\begin{table}[htb!]
    \centering
    \renewcommand{\arraystretch}{1.5} 
    \begin{tabular}{c|c}
        \hline
        Interactions & Vertices \\
        \hline\hline
        $\mathscr{G}\,H(p_2)\,H(p_3)$ & $(i\,\kappa/2)\,\left(p_{1\mu} p_{2\nu}+ p_{1\nu} p_{2\mu}- \eta_{\mu\nu}\, p_1\cdot p_2- \eta_{\mu\nu}\, m_h^2\right)$ \\
        \hline
        $\mathscr{G}\,V(p_1)\,V(p_2)$ & $-\,(i\kappa/2)\,m_V^2\,\left(\eta_{\alpha\mu}\,\eta_{\beta\nu}+\eta_{\alpha\nu}\,\eta_{\beta\mu}-\eta_{\alpha\beta}\,\eta_{\mu\nu}\right)+(i\kappa/2)\,\mathcal{C}_{\mu\nu,\alpha\beta}$
        \\
        \hline        $\mathscr{G}\,f(p_1)\,\bar{f}(p_2)$ & $(i\kappa/8)\,\left[\gamma_\mu (p_1 - p_2)_\nu+\gamma_\nu (p_1 - p_2)_\mu- 2\,\eta_{\mu\nu}\left(\slashed{p}_1- \slashed{p}_2\right)+ 4m_f\,\eta_{\mu\nu}\right]$\\
        \hline        $\mathscr{G}\,A(p_1)\,A(p_2)$ & $-(i\kappa/2)\,\left[\mathcal{C}_{\mu\nu,\alpha\beta}(p_1,p_2)- (p_1\cdot p_2)\,\mathcal{D}_{\mu\nu,\alpha\beta}\right]$ \\
        \hline
        $\mathscr{G}\,G(p_1)\,G (p_2)$ & $(i\kappa/2)\,\delta^{a_1 a_2}\,\left[
\mathcal{C}_{\mu\nu,\alpha\beta}(p_1,p_2)- (p_1\cdot p_2)\,\mathcal{D}_{\mu\nu,\alpha\beta}\right]$ \\
        \hline
    \end{tabular}
    \caption{3-legged vertices for the graviton-matter interactions in unitary gauge. Here, $H$ is the SM Higgs field, $V\in W^\pm,\,Z$ corresponds to the massive gauge bosons, $f$ corresponds to SM fermions and $G$ is the SM gluon field.}
    \label{tab:vertex}
\end{table}
\subsection{To Standard Model}
\subsubsection{Before EWSB}
\begin{align}\label{eq:grav-decay1}
& \Gamma_{\mathscr{G}\to \text{SM}\,\text{SM}}=\frac{m_{G_n}^3}{960\pi\,\Lambda_r^2}\,
\begin{dcases}
1\,, & \text{Higgs}\,,
\\[10pt]
6N_c\,, & \text{fermions}\,,
\\[10pt]
949\,, & \text{massless gauge bosons}\,,
\end{dcases}
\end{align}
where in the last line we have added decay into all massless gauge bosons, i.e., $B_\mu,\,W_\mu^i$ and gluons.
\subsubsection{After EWSB}
\begin{align}\label{eq:grav-decay2}
& \Gamma_{\mathscr{G}\to \text{SM}\,\text{SM}}=\frac{m_{G_n}^3}{960\pi\,\Lambda_r^2}\,
\begin{dcases}
\left(1-4x^2\right)^{5/2}\,, & \text{Higgs}\,,
\\[10pt]
6N_c\,\left(1-4r^2\right)^{3/2}\,\left(3+8r^2\right)\,, & \text{fermions}\,,
\\[10pt]
\delta\,\left(1-4r^2\right)^{1/2}\,\left(13+56\,r^2+48\,r^4\right)\,, & \text{massive gauge bosons}\,,
\\[10pt]
14 & \text{photons}\,,
\end{dcases}
\end{align}
where $r=m/m_{G_n}$ is the ratio of SM to graviton mass, and $\delta=1(2)$ for $Z\,(W^\pm)$ final states.
\subsection{To dark matter}
\begin{align}
&\Gamma_{\mathscr{G}\to\text{DM}\,\text{DM}}=\frac{m_{G_n}^3}{960\pi\,\Lambda_r^2}
\begin{dcases}
\left(1-4r^2\right)^{5/2}\,, & \text{spin-0}\,,
\\[10pt]
\left(1-4r^2\right)^{3/2}\,\left(3+8r^2\right)\,, & \text{spin-1/2}\,,
\\[10pt]
\left(1-4r^2\right)^{1/2}\,\left(13+56r^2+48r^4\right) & \text{spin-1}\,,
\end{dcases}
\end{align}
where $r=\mdm/m_{G_n}$.
\section{Graviton-mediated scattering cross-sections}
\label{sec:grav-med-cs}
In the following we consider, $x=2\,\mdm/\sqrt{s}$ and $y=2\,m_{\rm SM}/\sqrt{s}$, implying $y=0$ before EWSB.
\subsection{Before EWSB}
\subsubsection{Spin-0 DM}
\begin{align}
& \sigma_{\ell\ell\to SS}=\frac{s^3}{7680\pi\Lambda_r^4}\frac{(1-x^2)^{5/2}}{\left(s-m_{G_n}^2\right)^2+\Gamma_n^2\,m_{G_n}^2}\,,
\\&
\sigma_{qq\to SS}=\frac{s^3}{23040\pi\Lambda_r^4}\frac{(1-x^2)^{5/2}}{\left(s-m_{G_n}^2\right)^2+\Gamma_n^2\,m_{G_n}^2}\,,
\\&
\sigma_{VV\to SS}=\frac{13\,s^3}{51840\pi\,\Lambda_r^4}\frac{(1-x^2)^{5/2}}{\left(s-m_{G_n}^2\right)^2+\Gamma_n^2\,m_{G_n}^2}\,,
\\&
\sigma_{hh\to SS}=\frac{s^3}{5760\pi \Lambda_r^4}\frac{(1-x^2)^{5/2}}{\left(s-m_{G_n}^2\right)^2+\Gamma_n^2\,m_{G_n}^2}\,,
\\&
\sigma_{gg\to SS}=\frac{7s^3}{92160\pi\Lambda_r^4}\frac{(1-x^2)^{5/2}}{\left(s-m_{G_n}^2\right)^2+\Gamma_n^2\,m_{G_n}^2}\,,
\\&
\sigma_{\gamma\gamma\to SS}=\frac{7s^3}{11520\pi\Lambda_r^4}\frac{(1-x^2)^{5/2}}{\left(s-m_{G_n}^2\right)^2+\Gamma_n^2\,m_{G_n}^2}\,.
\end{align}
\subsubsection{Spin-1 DM}
\begin{align}
& \sigma_{\ell\ell\to XX}=\frac{s^3}{7680\,\pi\Lambda_r^4}\,\frac{\left(3\,x^4+14\,x^2+13\right)}{\left(s-m_{G_n}^2\right)^2+\Gamma_n^2\,m_{G_n}^2}\,\sqrt{1-x^2}\,,
\\&
\sigma_{qq\to XX}=\frac{s^3}{23040\,\pi\Lambda_r^4}\,\frac{\left(3\,x^4+14\,x^2+13\right)}{\left(s-m_{G_n}^2\right)^2+\Gamma_n^2\,m_{G_n}^2}\,\sqrt{1-x^2}\,,
\\&
\sigma_{VV\to XX}=\frac{s^3}{25920\pi\,\Lambda_r^4}\,\frac{\left(3 x^4+14 x^2+13\right)}{\left(s-m_{G_n}^2\right)^2+\Gamma_n^2\,m_{G_n}^2}\,\sqrt{1-x^2}\,,
\\&
\sigma_{hh\to XX}=\frac{s^3}{5760\,\pi\Lambda_r^4}\,\frac{\left(3\,x^4+14\,x^2+13\right)}{\left(s-m_{G_n}^2\right)^2+\Gamma_n^2\,m_{G_n}^2}\,\sqrt{1-x^2}\,,
\\&
\sigma_{gg\to XX}=\frac{7s^3}{92160\,\pi\Lambda_r^4}\,\frac{\left(3\,x^4+14\,x^2+13\right)}{\left(s-m_{G_n}^2\right)^2+\Gamma_n^2\,m_{G_n}^2}\,\sqrt{1-x^2}\,,
\\&
\sigma_{\gamma\gamma\to XX}=\frac{7s^3}{11520\,\pi\Lambda_r^4}\,\frac{\left(3\,x^4+14\,x^2+13\right)}{\left(s-m_{G_n}^2\right)^2+\Gamma_n^2\,m_{G_n}^2}\,\sqrt{1-x^2}\,.
\end{align}
\subsubsection{Spin-1/2 DM}
\begin{align}
&\sigma_{\ell\ell\to NN}=\frac{s^3}{7680\pi\,\Lambda_r^4}\,\frac{2x^2+3}{\left(s-m_{G_n}^2\right)^2+\Gamma_n^2\,m_{G_n}^2}\,\left(1-x^2\right)^{3/2}\,,
\\&
\sigma_{qq\to NN}=\frac{s^3}{23040\pi\,\Lambda_r^4}\,\frac{2x^2+3}{\left(s-m_{G_n}^2\right)^2+\Gamma_n^2\,m_{G_n}^2}\,\left(1-x^2\right)^{3/2}\,,
\\&
\sigma_{VV\to NN}=\frac{13s^3}{51840\pi\,\Lambda_r^4}\,\frac{2x^2+3}{\left(s-m_{G_n}^2\right)^2+\Gamma_n^2\,m_{G_n}^2}\,\left(1-x^2\right)^{3/2}\,,
\\&
\sigma_{hh\to NN}=\frac{s^3}{5760\pi\,\Lambda_r^4}\,\frac{2x^2+3}{\left(s-m_{G_n}^2\right)^2+\Gamma_n^2\,m_{G_n}^2}\,\left(1-x^2\right)^{3/2}\,,
\\&
\sigma_{gg\to NN}=\frac{7s^3}{92160\pi\,\Lambda_r^4}\,\frac{3-2x^4-x^2}{\left(s-m_{G_n}^2\right)^2+\Gamma_n^2\,m_{G_n}^2}\,\left(1-x^2\right)^{1/2}\,,
\\&
\sigma_{\gamma\gamma\to NN}=\frac{7s^3}{11520\pi\,\Lambda_r^4}\,\frac{3-2x^4-x^2}{\left(s-m_{G_n}^2\right)^2+\Gamma_n^2\,m_{G_n}^2}\,\left(1-x^2\right)^{1/2}\,.
\end{align}
\subsection{After EWSB}
\subsubsection{Spin-0 DM}
\begin{align}
& \sigma_{\ell\ell\to SS}=\frac{s^3}{23040\pi\,\Lambda_r^4}\,\frac{(1-x^2)^{5/2}\,(2y^2+3)}{\left(s-m_{G_n}^2\right)^2+\Gamma_n^2\,m_{G_n}^2}\,\sqrt{1-y^2}\,,
\\&
\sigma_{qq\to SS}=\frac{s^3}{69120\pi\,\Lambda_r^4}\,\frac{(1-x^2)^{5/2}\,(2y^2+3)}{\left(s-m_{G_n}^2\right)^2+\Gamma_n^2\,m_{G_n}^2}\,\sqrt{1-y^2}\,,
\\&
\sigma_{VV\to SS}=\frac{s^3}{51840\pi\,\Lambda_r^4}\,\frac{(3y^4+14y^2+13)\,(1-x^2)^2}{\left(s-m_{G_n}^2\right)^2+\Gamma_n^2\,m_{G_n}^2}\,\sqrt{\frac{1-x^2}{1-y^2}}\,,
\\&
\sigma_{hh\to SS}=\frac{s^3}{5760\pi\,\Lambda_r^4}\,\frac{(1-x^2)^{5/2}\,(1-y^2)^{3/2}}{\left(s-m_{G_n}^2\right)^2+\Gamma_n^2\,m_{G_n}^2}\,.
\end{align}
\subsubsection{Spin-1 DM}
\begin{align}
& \sigma_{\ell\ell\to XX}=\frac{s^3}{23040\pi\,\Lambda_r^4}\,\frac{\left(3x^4+14x^2+13\right)\,\left(2y^2+3\right)}{\left(s-m_{G_n}^2\right)^2+\Gamma_n^2\,m_{G_n}^2}\,\sqrt{(1-x^2)\,(1-y^2)}\,,
\\&
\sigma_{qq\to XX}=\frac{s^3}{69120\pi\,\Lambda_r^4}\,\frac{\left(3x^4+14x^2+13\right)\,\left(2y^2+3\right)}{\left(s-m_{G_n}^2\right)^2+\Gamma_n^2\,m_{G_n}^2}\,\sqrt{(1-x^2)\,(1-y^2)}\,,
\\&
\sigma_{VV\to XX}=\frac{s^3}{25920\pi\,\Lambda_r^4}\,\frac{\left(3 x^4+14 x^2+13\right)\,\left(3 y^4+14 y^2+13\right)}{\left(s-m_{G_n}^2\right)^2+\Gamma_n^2\,m_{G_n}^2}\,\sqrt{\frac{1-x^2}{1-y^2}}\,,
\\&
\sigma_{hh\to XX}=\frac{s^3}{5760\pi\,\Lambda_r^4}\,\frac{\left(3x^4+14x^2+13\right)}{\left(s-m_{G_n}^2\right)^2+\Gamma_n^2\,m_{G_n}^2}\,\sqrt{(1-x^2)}\,(1-y^2)^{3/2}\,.
\end{align}
\subsubsection{Spin-1/2 DM}
\begin{align}
&\sigma_{\ell\ell\to NN}=\frac{s^3}{23040\pi\,\Lambda_r^4}\,\frac{(2x^2+3)\,(2y^2+3)}{\left(s-m_{G_n}^2\right)^2+\Gamma_n^2\,m_{G_n}^2}\,\sqrt{1-y^2}\,(1-x^2)^{3/2}\,,
\\&
\sigma_{qq\to NN}=\frac{s^3}{69120\pi\,\Lambda_r^4}\,\frac{(2x^2+3)\,(2y^2+3)}{\left(s-m_{G_n}^2\right)^2+\Gamma_n^2\,m_{G_n}^2}\,\sqrt{1-y^2}\,(1-x^2)^{3/2}\,,
\\&
\sigma_{VV\to NN}=\frac{s^3}{51840\pi\,\Lambda_r^4}\,\frac{(2x^2+3)\,\left(13 + 14 y^2 + 3 y^4\right)}{\left(s-m_{G_n}^2\right)^2+\Gamma_n^2\,m_{G_n}^2}\,\frac{(1-x^2)^{3/2}}{\sqrt{1-y^2}}\,,
\\&
\sigma_{hh\to NN}=\frac{s^3}{5760\pi\,\Lambda_r^4}\,\frac{2x^2+3}{\left(s-m_{G_n}^2\right)^2+\Gamma_n^2\,m_{G_n}^2}\,(1-x^2)^{3/2}\,(1-y^2)^{3/2}\,.
\end{align}
\twocolumngrid
%

\end{document}